\documentclass[11pt]{article}
\usepackage{amsthm,amsfonts,%showkeys
  }
\swapnumbers
\font\notefont=cmsl8

\theoremstyle{plain}
\newtheorem{thm}{THEOREM}[section]
\newtheorem{lm}[thm]{LEMMA}
\newtheorem{cl}[thm]{COROLLARY}

\theoremstyle{definition}

\theoremstyle{remark}

\newcommand{\upchi}{\raise1pt\hbox{$\chi$}}
\newcommand{\R}{{\mathord{\mathbb R}}}
\newcommand{\C}{{\mathord{\mathbb C}}}
\newcommand{\Z}{{\mathord{\mathbb Z}}}
\newcommand{\N}{{\mathord{\mathbb N}}}
\newcommand{\supp}{{\mathop{\rm supp\ }}}
\newcommand{\mfr}[2]{{\textstyle\frac{#1}{#2}}}
\newcommand{\const}{C}
\newcommand{\iint}{\mathop{\displaystyle\int\!\!\int}}
\newcommand{\hw}{{\mathord{\widehat{w}}^{\phantom{*}}}}
\newcommand{\hn}{{\mathord{\widehat{n}}}}
\newcommand{\an}{{\mathord{a}}^{\phantom{*}}}
\newcommand{\bn}{{\mathord{b}}^{\phantom{*}}}
\newcommand{\atn}{{\mathord{\widetilde{a}}}^{\phantom{*}}}
\newcommand{\at}{{\mathord{\widetilde{a}}}}
\newcommand{\cA}{{\mathord{\cal A}}}
\newcommand{\cB}{{\mathord{\cal B}}}
\begin{document}
\title{GROUND STATE ENERGY OF THE ONE-COMPONENT CHARGED BOSE GAS}
\author{
  \begin{tabular}{ccc}
    Elliott H. Lieb\thanks{Work partially supported by U.S. National Science
      Foundation grant PHY98 20650-A01.} &\hspace{2cm}& Jan Philip
    Solovej\thanks{Work partially supported by EU TMR grant, by the Danish 
      research foundation center MaPhySto, and by a grant from the Danish research
      council  \newline \hfill \copyright 2000 \ by the authors. This article may be
      reproduced in its entirety for non-commercial purposes. }\\
    \normalsize Departments of Physics and Mathematics && 
    \normalsize Department of Mathematics\\ 
    \normalsize Jadwin Hall, Princeton University && \normalsize University of Copenhagen\\
    \normalsize PO Box 708 &&\normalsize Universitetsparken 5\\
    \normalsize  Princeton, N.J. 08544-0708 & &
    \normalsize DK-2100 Copenhagen, Denmark\\
    \normalsize {\it e-mail\/}: lieb@princeton.edu &&
    \normalsize {\it e-mail\/}: solovej@math.ku.dk
  \end{tabular}
  \bigskip
  \date{August 8, 2000 \\ 
    $ {\phantom x}  $  \\
 Dedicated to Leslie L. Foldy on the occasion of his 80th birthday }}

                                \maketitle

\begin{abstract}
  The model considered here is the `jellium' model in which there is
  a uniform, fixed background with charge density $-e\rho$ in a large
  volume $V$ and in which $N=\rho V$ particles of electric charge $+e$
  and mass $m$ move --- the whole system being neutral.  In 1961 Foldy
  used Bogolubov's 1947 method to investigate the ground state energy of
  this system for bosonic particles in the large $\rho$ limit. He found
  that the energy per particle is $-0.402 \, r_s^{-3/4} {me^4}/{\hbar^2}$
  in this limit, where $r_s=(3/4\pi \rho)^{1/3}e^2m/\hbar^2$.  Here we
  prove that this formula is correct, thereby validating, for the first
  time, at least one aspect of Bogolubov's pairing theory of the Bose gas.
\end{abstract}

\vfill\eject
\section{Introduction}

Bogolubov's 1947 pairing theory \cite{B} for a Bose fluid was used
by Foldy \cite{F} in 1961 to calculate the ground state energy of the
one-component plasma (also known as ``jellium") in the high density
regime --- which is the regime where the Bogolubov method was thought
to be exact for this problem.  Foldy's result will be verified
rigorously in this paper; to our knowledge, this is the first example
of such a verification of Bogolubov's theory in a
three-dimensional system of bosonic particles.

Bogolubov proposed  his approximate theory of the Bose fluid \cite{B} in an
attempt to explain the properties of liquid Helium. His main contribution
was the concept of pairing of particles with momenta $k$ and $-k$; these
pairs are supposed to be the basic constituents of the ground state
(apart from the macroscopic fraction of particles in the ``condensate", or
$k=0$ state) and they are the basic unit of the elementary
excitations of the system.  The pairing concept was later generalized to
fermions, in which case the pairing was between particles having 
opposite momenta and, at the same time, opposite spin. Unfortunately, this
appealing concept about the boson ground state has neither been
verified rigorously in a 3-dimensional example, nor has it been
conclusively verified experimentally (but pairing has been verified
experimentally for superconducing electrons).  

The simplest question that can be asked is the correctness of the
prediction for the ground state energy (GSE). This, of course, can only be
exact in a certain limit -- the `weak coupling' limit. In the case
of the charged Bose gas, interacting via Coulomb forces, this corresponds
to the {\it high density} limit. In gases with short range forces
the weak coupling limit corresponds to low density instead.

Our system has $N$ bosonic particles with unit positive charge and
coordinates $x_j$, and a uniformly negatively charged `background' in a
large domain $\Omega$ of volume $V$. We are interested in the
thermodynamic limit. A physical realization of this model is supposed to
be a uniform electron sea in a solid, which forms the background, while
the moveable  `particles' are bosonic atomic nuclei. The particle number
density is then $\rho = N/V$ and this number is also the charge density
of the background, thus ensuring charge neutrality.

The Hamiltonian of the one-component plasma is
\begin{equation}\label{ham} 
  H= \frac{1}{2} \sum_{j=1}^N p_j^2 + U_{pp} +U_{pb} + U_{bb}
\end{equation}
where ${ p}=-i{ \nabla}$ is the  momentum operator, $p^2 =-\Delta$, 
and the three potential energies, particle-particle, particle-background
and background-background, are given by
\begin{eqnarray}
  U_{pp}&=&\sum_{1\leq i < j \leq N} |x_i-x_j|^{-1}  \label{pot1}\\
  U_{pb}&=&-\rho\sum_{j=1}^N  \int_{\Omega} |x_j-y|^{-1}\,d^3y \label{pot2}\\
  U_{bb}&=&\mfr{1}{2}\rho^2\int_{\Omega} \int_{\Omega}
  |x-y|^{-1}\,d^3xd^3y \ . 
  \label{pot3}
\end{eqnarray}

In our units $\hbar^2/m =1$ and the charge is $e=1$. The `natural' energy
unit we use is two Rydbergs, $2Ry = me^4/\hbar^2$. It is customary to
introduce the dimensionless quantity $r_s=(3/4\pi \rho)^{1/3}e^2m/\hbar^2$.
High density is small $r_s$.

The Coulomb  potential is infinitely  long-ranged  and great care has
to be taken because the finiteness of the energy per particle in the
thermodynamic limit depends, ultimately, on delicate cancellations.  The
existence of the thermodynamic limit for a system of positive and negative
particles, with the negative ones being fermions, was shown only in 1972
\cite{LLe} (for the free energy, but the same proof works for the ground
state energy). Oddly, the jellium case is technically a bit harder, and this
was done in 1976 \cite{LN} (for both bosons and fermions). One conclusion
from this work is that neutrality (in the thermodynamic limit)  will come
about automatically -- even if one does not assume it -- provided one allows
any excess charge to escape to infinity. In other words, given the background
charge, the choice of a neutral number of particles has the lowest energy in
the thermodynamic limit. A second point, as shown in \cite{LN}, is that $e_0$
is independent of the shape of the domain $\Omega$ provided the boundary is
not too wild. For Coulomb systems this is not trivial and for real magnetic
systems it is not even generally true.  We take advantage of this liberty and
assume that our domain is a cube $[0,L]\times [0,L]\times [0,L]$ with $L^3=V$.

We note the well-known fact that the lowest energy of $H$ in
(\ref{ham}) without any restriction about `statistics' (i.e., on the
whole of $\otimes^N L^2(\R^{3})$) is the same as for bosons, i.e., on
the symmetric subspace of $\otimes^N L^2(\R^{3})$.
The fact that bosons  have the lowest energy comes from the 
Perron-Frobenius Theorem applied to $-\Delta$.

Foldy's calculation leads to the following theorem about the 
asymptotics of the energy for small $r_s$, which we call
Foldy's law.

\begin{thm}[Foldy's Law]\label{thm:foldy}\hfill\\
Let $E_0$ denote the ground state energy, 
i.e., the bottom of the
spectrum, of the Hamiltonian $H$
acting in the Hilbert space $\otimes^N L^2(\R^{3})$. We assume that 
$\Omega=[0,L]\times [0,L]\times [0,L]$.
The ground state energy  per particle, $e_0=E_0/N$ , in the thermodynamic
limit $N,L\to\infty$ with $N/V=\rho$ fixed, in units of ${m
  e^4}/\hbar^2$, is 
\begin{equation} 
  \label{gse} 
  \lim_{V\to \infty} E_0/N  =e_0 =  -0.40154\, r_s^{-3/4} + o(\rho^{1/4})=
  -0.40154 \left(\frac{4\pi}{3}\right)^{1/4} \rho^{1/4} + o(\rho^{1/4}) \ ,
\end{equation}
where the number -0.40154 is, in fact, the integral 
\begin{equation} \label{int} 
  A= \frac{1}{\pi} 6^{1/4} \int_0^{\infty} \left\{p^2(p^4+2)^{1/2}
    -p^4 -1 \right\} dp =-\frac{
    3^{1/4}4\,\Gamma(3/4)}{5\sqrt{\pi}\,\Gamma(5/4)}
\approx -0.40154\ .  
\end{equation}
\end{thm}

Actually, our proof gives a result that is more general than Theorem
\ref{thm:foldy}. We allow the  particle number $N$ to be totally arbitrary,
i.e., we do not require $N=\rho V$. Our lower bound is still given by
(\ref{gse}), where now $\rho$ refers to the background charge density.

In \cite{F} 0.40154 is replaced by 0.80307 since the energy unit there is 1
Ry.  The main result of our paper is to prove (\ref{gse}) by
obtaining a lower bound on $E_0$ that agrees with the right side of 
(\ref{gse}) An upper bound to $E_0$ that agrees with (\ref{gse}) (to
leading order) was given in 1962 by Girardeau \cite{GM}, using the
variational method of himself and Arnowitt \cite{GA}.  Therefore, to verify
(\ref{gse}) to leading order it is only necessary to construct a rigorous
lower bound of this form and this will be done here. It has to be
admitted, as explained below, that the problem that Foldy and Girardeau treat
is slightly different from ours because of different boundary conditions and
a concommitant different treatment of the background. We regard this
difference as a technicality that should be cleared up one day, and do not
hesitate to refer to the statement of \ref{thm:foldy} as a theorem.

Before giving our proof, let us remark on a few historical and conceptual
points. Some of the early history about the Bose gas, can be found in the
lecture notes \cite{L}.

Bogolubov's analysis starts by assuming periodic boundary condition on the
big box $\Omega$ and writing everything in momentum (i.e., Fourier) space. 
The values of the momentum, $k$ are then discrete: \ $k=(2\pi/L) (m_1, m_2,
m_3)$ with $m_i$ an integer.  A convenient tool for taking care of various
$n!$ factors is to introduce second quantized operators $a_k^{\#}$ (where
$a^{\#}$ denotes $a$ or $a^*$), but it has to be understood that this is
only a bookkeeping device. Almost all authors worked in momentum space, but
this is neither necessary nor necessarily the most convenient
representation (given that the calculations are not rigorous). Indeed,
Foldy's result was reproduced by a calculation entirely in $x-$space
\cite{LS}.  Periodic boundary conditions are not physical, but that was
always chosen for convenience in momentum space.  

We shall instead let the particle move in the whole space, i.e., the operator
$H$ acts in the Hilbert space $L^2(\R^{3N})$, or rather, since we consider
bosons, in the the subspace consisting of the $N$-fold fully symmetric tensor
product of $L^2(\R^{3})$. The background potential defined in (\ref{pot1}) is
however still localized in the cube $\Omega$.  We could also have confined
the particles to $\Omega$ with Dirichlet boundary conditions. This would only
raise the ground state energy and thus, for the lower bound, our setup is
more general.

There is, however, a technical point that has to be considered when dealing
with Coulomb forces. The background never appears in Foldy's calculation; he
simply removes the $k=0$ mode from the Fourier transform, $\nu$ of the
Coulomb potential (which is $\nu (k) = 4\pi |k|^{-2}$, but with $k$ taking
the discrete values mentioned above, so that we are thus dealing with a
`periodized' Coulomb potential). The $k=0$ elimination means that we set
$\nu(0) =0$, and this amounts to a subtraction of the average value of the
potential -- which is supposed to be a substitute for the effect of a
neutralizing background.  It does not seem to be a trivial matter to
prove that this is equivalent to having a background, but it surely can be
done. Since we do not wish to overload this paper, we leave this
demonstration to another day.  In any case the answers agree (in the sense
that our rigorous lower bound agrees with Foldy's answer), as we prove here.
If one accepts the idea that setting $\nu(0) =0$ is equivalent to having
a neutralizing background, then the ground state energy problem is finished
because Girardeau shows \cite{GM} that Foldy's result is a true upper bound
within the context of the $\nu(0) =0$ problem.

The potential energy is quartic in the operators $a_k^\#$. In
Bogolubov's analysis only terms in
which there are four or two $a_0^\#$ operators are retained. The operator
$a_0^*$ creates, and  $a_0$ destroys particles with momentum $0$ and such
particles are the constituents of the `condensate'.  In general there are no
terms with three $a_0^\#$ operators (by momentum conservation) and in Foldy's
case there is also no four $a_0^\#$ term (because of the subtraction just
mentioned).  

For the usual short range potential there is a four $a_0^\#$  term and this
is supposed to give the leading term in the energy, namely $e_0= 4\pi \rho
a$, where $a$ is the `scattering length' of the two-body potential.  Contrary
to what would seem reasonable, this number, $4\pi \rho a$ is {\it not} the
coefficient of the four $a_0^\#$  term, and to to prove that $4\pi \rho a$
is, indeed, correct took some time. It was done in 1998 \cite{LY} and the
method employed in \cite{LY} will play an essential role here. But it is
important to be clear about the fact that the four $a_0^\#$, or `mean field'
term is absent in the jellium case by virtue of charge neutrality.  The
leading term in this case  presumably comes from the two $a_0^\#$ terms, and
this is what we have to prove. For the short range case, on the other hand,
it is already difficult enough to obtain the $4\pi \rho a$ energy that going
beyond this to the two $a_0^\#$ terms is beyond the reach of rigorous
analysis at the moment.

The Bogolubov ansatz presupposes the existence of Bose Einstein condensation
(BEC). That is, most of the particles are in the $k=0$ mode and the few that
are not come in pairs with momenta $k$ and $-k$. Two things must be said
about this. One is that the only case (known to us) in which one can verify
the correctness of the Bogolubov picture at weak coupling is the {\it
one-dimensional} delta-function gas \cite{LLi} --- in which case there is
presumably {\it no} BEC (because of the low dimensionality). Nevertheless the
Bogolubov picture remains correct at low density and the explanation of this
seeming contradiction lies in the fact that BEC is not needed; what is really
needed is a kind of condensation on a length scale that is long compared to
relevant parameters,  but which is fixed and need not be as large as the box
length $L$.  This was realized in \cite{LY} and the main idea there was to
decompose $\Omega$ into fixed-size boxes of appropriate length and use
Neumann boundary conditions on these boxes (which can only lower the energy,
and which is fine since we want a lower bound).  We shall make a similar
decomposition here, but, unlike the case in \cite{LY} where the potential is
purely repulsive, we must deal here  with the Coulomb potential and work hard
to achieve the necessary cancellation.

The only case in which BEC has been proved to exist is in the hard core
lattice gas  at half-filling (equivalent to the spin-1/2 $XY$ model)
\cite{KLS}.  

Weak coupling is sometimes said to be a `perturbation theory' regime, but
this is not really so. In the one-dimensional case \cite{LLi} the asymptotics
near $\rho=0$ is extremely difficult to deduce from the exact solution
because the `perturbation' is singular. Nevertheless, the Bogolubov
calculation gives it effortlessly, and this remains a mystery.

One way to get an excessively negative lower bound to $e_0$ for jellium is to
ignore the kinetic energy. One can then show easily (by an argument due to
Onsager) that the potential energy alone is bounded below by $e_0 \sim
-\rho^{1/3}$. See \cite{LN}. Thus, our goal is to show that the kinetic
energy raises the energy to $-\rho^{1/4}$.  This was done, in fact, in
\cite{CLY}, but without achieving the correct coefficient $-0.803
(4\pi/3)^{1/4}$. Oddly, the $-\rho^{1/4}$ law was proved in \cite{CLY} by
first showing that the {\it non-thermodynamic} $N^{7/5}$ law for a {\it
two-component} bosonic plasma, as conjectured by Dyson \cite{D}, is correct.

The \cite{CLY} paper contains an important innovation that will play a key
role here. There, too, it was necessary to decompose ${\bf R}^3$ into boxes,
but a way had to be found to eliminate the Coulomb interaction {\it between}
different boxes. This was accomplished by {\it not} fixing the location of
the boxes but rather averaging over all possible locations of the boxes. This
`sliding localization' will play a key role here, too. This idea was expanded
upon in \cite{GG}.  Thus, we shall have to consider only one finite box with
the particles and the background charge in it independent of the rest of the
system.  However, a price will have to be paid for this luxury, namely it
will not be entirely obvious that the number of particles we want to place in
each box is the same for all boxes, i.e., $\rho \ell^3$, where $\ell$ is the
length of box.  Local neutrality, in other words, cannot be taken for
granted. The analogous problem in \cite{LY} is easier because no attractive
potentials are present there. We solve this problem by choosing the
number, $n$, in each box to be the number that gives the lowest energy in
the box. This turns out to be close to $n=\rho \ell^3$, as we show and as we
know  from \cite{LN} must be the case as $\ell \to \infty$.

Finally, let us remark on one bit of dimensional analysis that the reader
should keep in mind. One should not conclude from (\ref{gse}) that a typical
particle has energy $\rho^{1/4}$ and hence momentum $\rho^{1/8}$ or de
Broglie wavelength $\rho^{-1/8}$. This is {\it not} the correct picture. 
Rather, a glance at the Bogolubov/Foldy calculation shows that the momenta of
importance are of order $\rho^{-1/4}$, and the seeming paradox is resolved by
noting that the number of excited particles (i.e., those not in the $k=0$
condensate) is of order $N\rho^{-1/4}$.  This means that we can, hopefully,
localize particles to lengths as small as $\rho^{-1/4 + \epsilon}$, and cut
off the Coulomb potential at similar lengths, without damage, provided we do
not disturb the condensate particles. It is this clear separation of scales
that enables our asymptotic analysis to succeed.

\section{Outline of the Proof}

The proof of our main theorem~\ref{thm:foldy} is rather complicated and
somewhat hard to penetrate, so we present the following outline to
guide the reader.

\subsection{Section~\ref{sec:sliding}}
Here we localize the system whose size is $L$ into small boxes of size
$\ell$ independent of $L$, but dependent on the intensive quantity
$\rho$.  Neumann boundary conditions for the Laplacian are used in
order to ensure a lower bound to the energy. 
We always think of operators in terms of quadratic forms and the
Neumann Laplacian in a box $Q$ is defined for all functions  in 
$\psi\in L^2(Q)$ by the quadratic form
$$
 (\psi,-\Delta_{\rm Neumann}\psi)=\int_Q|\nabla\psi(x)|^2\,dx. 
$$
The lowest eigenfunction
of the Neumann Laplacian is the constant function and this plays the
role of the condensate state. This state not only minimizes the,
kinetic energy, but it is also consistent with neutralizing the
background and thereby minimizing the Coulomb energy.  The particles
not in the condensate will be called `excited' particles.

To avoid localization errors we take $\ell\gg\rho^{-1/4}$, which is
the relevant scale as we mentioned in the Introduction.  The
interaction among the boxes is controlled by using the sliding method
of \cite{CLY}. The result is that we have to consider only
interactions among the particles and the background in each little box
separately.

The $N$ particles have to be distributed among the boxes in a way that
minimizes the total energy. We can therefore not assume that each box
is neutral. Instead of dealing with this distribution problem we do a
simpler thing which is to choose the particle number in each little
box so as to achieve the absolute minimum of the energy in that box.
Since all boxes are equivalent this means that we take a common value
$n$ as the particle number in each box. The total particle number
which is $n$ times the number of boxes will not necessarily equal $N$,
but this is of no consequence for a lower bound.  We shall show later,
however, that it equality is nearly achieved, i.e., the the energy
minimizing number $n$ in each box is close to the value needed for
neutrality.

\subsection{Section~\ref{sec:cutoff}}
It will be important for us to replace the Coulomb potential by a
cutoff Coulomb potential. There will be a short distance cutoff of the
singularity at a distance $r$ and a large distance cutoff of the tail
at a distance $R$, with $ r\leq R\ll\ell $.  One of the unusual
features of our proof is that $r$ are $R$ are not fixed once and for
all, but are readjusted each time new information is gained about the
error bounds.

In fact, already in Sect.~\ref{sec:cutoff} we give a simple
preliminary bound on $n$ by choosing $R\sim\rho^{-1/3}$, which is much
smaller than the relevant scale $\rho^{-1/4}$, although the choice of
$R$ that we shall use at the end of the proof is of course much larger
than $\rho^{-1/4}$, but less than $\ell$.

\subsection{Section~\ref{sec:estimates}}
There are several terms in the Hamiltonian. There is the kinetic
energy, which is non-zero only for the excited particles.  The
potential energy, which is a quartic term in the language of second
quantization, has various terms according to the number of times the
constant function appears. Since we do not have periodic boundary
conditions we will not have the usual simplification caused by
conservation of momentum,  and the potential energy will be
correspondingly more complicated than the usual expression found in
textbooks.

In this section we give bounds on the different terms in the
Hamiltonian and use these to get a first control on the condensation,
i.e., a control on the number of particles $\hn_+$ in each little box
that are not in the condensate state.

The difficult point is that $\hn_+$ is an operator that does not
commute with the Hamiltonian and so it does not have a sharp value in
the ground state.  We give a simple preliminary bound on its average
$\langle \hn_+\rangle$ in the ground state by again choosing
$R\sim\rho^{-1/3}$.  In order to control the condensation to an
appropriate accuracy we shall eventually need not only a bound on the
average, $\langle \hn_+\rangle$, but also on the fluctuation, i.e, on
$\langle \hn_+^2\rangle$. This will be done in Sect.~\ref{sec:local}
using a novel method developed in Appendix~\ref{app:local} for
localizing off-diagonal matrices.

\subsection{Section~\ref{sec:quadratic}}
The part of the potential energy that is most important is the part
that is quadratic in the condensate operators $a_0^\#$ and
quadratic in the excited variables $a_p^\#$ with $p\ne0$.  This,
together with the kinetic energy, which is also quadratic in the
$a_p^\#$, is the part of the Hamiltonian that leads to Foldy's
law.  Although we have not yet managed to eliminate the non-quadratic
part up to this point we study the main `quadratic' part of the
Hamiltonian.  It is in this section that we essentially do Foldy's
calculation.

It is not trivial to diagonalize the quadratic form and thereby
reproduce Foldy's answer because there is no momentum conservation.
In particular there is no simple relation between the resolvent of the
Neumann Laplacian and the Coulomb kernel.  The former is defined
relative to the box and the latter is defined relative to the whole of
$\R^3$.  It is therefore necessary for us to localize the wavefunction
in the little box away from the boundary. On such functions the
boundary condition is of no importance and we can identify the kinetic
energy with the Laplacian in all of $\R^3$. This allows us to have a
simple relation between the Coulomb term and the kinetic energy term
since the Coulomb kernel is in fact the resolvent of the Laplacian in
all of $\R^3$.

When we cut off the wavefunction near the boundary we have to be very
careful because we must not cut off the part corresponding to the
particles in the condensate. To do so would give too large a
localization energy. Rather, we cut off only functions with
sufficiently large kinetic energy so that the localization energy is
relatively small compared to the kinetic energy.  The technical lemma
needed for this is a double commutator inequality given in
Appendix~\ref{app:double}.

\subsection{Section~\ref{sec:nn+}}
At this point we have bounds available for the quadratic part (from
Sect.~\ref{sec:quadratic}) and the annoying non-quadratic part (from
Sect.~\ref{sec:estimates}) of the Hamiltonian. These depend on $r$,
$R$, $n$, $\langle \hn_+\rangle$, and $\langle \hn_+^2\rangle$.  We
avail ourselves of the bounds previously obtained for $n$ and $\langle
\hn_+\rangle$ and now use our freedom to choose different values for
$r$ and $R$ to bootstrap to the desired bounds on $n$ and $\langle
\hn_+\rangle$, i.e., we prove that there is almost neutrality and
almost condensation in each little box.

\subsection{Section~\ref{sec:local}}
In order to control $\langle \hn_+^2\rangle$ we utilize, for the first
time, the new method for localizing large matrices given in
Appendix~\ref{app:local}.  This method allows us to restrict to states
with small fluctuations in $\hn_+$, and thereby bound $\langle
\hn_+^2\rangle$, provided we know that the terms that do not commute
with $\hn_+$ have suffciently small expectation values.  We then give
bounds on these $\hn_+$ `off-diagonal' terms.  Unfortunately, these
bounds are in terms of positive quantities coming from the Coulomb
repulsion, but for which we actually do not
have independent a-priori bounds.  Normally, when proving a lower
bound to a Hamiltonian, we can sometimes control error terms by
absorbing them into positive terms in the Hamiltonian, which are then
ignored.  This may be done even when we do not have an a-priori bound
on these positive terms. If we want to use Theorem~\ref{local} in
Appendix~\ref{app:local}, we will need an absolute bound on the
`off-diagonal' terms and we can therefore not use the technique of
absorbing them into the positive terms.  The decision when to use
the theorem in  Appendix~\ref{app:local} or use the technique of
absorption into positive terms is resolved in
Section~\ref{sec:foldyslaw}.

\subsection{Section~\ref{sec:foldyslaw}}

Since we do not have an a-priori bound on the positive Coulomb terms
as described above we are faced with a dichotomy. If the positive
terms are, indeed, so large that enough terms can be controlled by
them we do not need to use the localization technique of
Appendix~\ref{app:local} to finish the proof of Foldy's law.  The second
possibility is that the positive terms  are bounded in which case we
can use this fact to control the terms that do commute with $\hn_+$
and this allows us to use the localization technique in
Appendix~\ref{app:local} to finish the proof of Foldy's law.  Thus,
the actual magnitude of the positive repulsion terms is unimportant
for the derivation of Foldy's law.

\section{Reduction to a small box}\label{sec:sliding}
As described in the previous sections
we shall localize the problem into smaller cubes of size $\ell\ll L$. 
We shall in fact choose $\ell$ as a function of $\rho$ in such a way
that $\rho^{1/4}\ell\to\infty$ as $\rho\to\infty$. 
  
We shall localize the kinetic energy by using Neumann boundary
conditions on the smaller boxes.  

We shall first, however, describe how we may control the electostatic
interaction between the smaller boxes using the sliding technique of
\cite{CLY}.

Let $t$, with $0<t<1/2$, be a parameter which we shall choose later to
depend on $\rho$ in such a way that $t\to0$ as $\rho\to\infty$.

The choice of $\ell$ and $t$ as functions of $\rho$ will be made at
the end of section~\ref{sec:foldyslaw} when we complete the proof of
Foldy's law.

Let $\upchi\in C^\infty_0(\R^3)$ satisfy $\supp\upchi\subset
\left[(-1+t)/2,(1-t)/2\right]^3$, $0\leq \upchi\leq1$, $\upchi(x)=1$
for $x$ in the smaller box $\left[(-1+2t)/2,(1-2t)/2\right]^3$, and
$\upchi(x)=\upchi(-x)$.  Assume that all $m$-th order derivatives of
$\upchi$ are bounded by $\const_m t^{-m}$, where the constants
$\const_m$ depend only on $m$ and are, in particular, independent of
$t$.  Let $\upchi_\ell(x)=\upchi(x/\ell)$.  Let
$\eta=\sqrt{1-\upchi}$. We shall assume that $\upchi$ is defined such
that $\eta$ is also $C^1$. Let $\eta_\ell(x)=\eta(x/\ell)$.
 Using $\upchi$ we define the constant $\gamma $ by  
$\gamma^{-1} = \int\chi(y)^2\,dy$, and note that $1\leq
\gamma\leq(1-2t)^{-3}$. 
We also introduce the Yukawa potential $Y_{\nu}(x)=|x|^{-1}e^{-\nu|x|}$ for
$\nu>0$.

As a preliminary to the following lemma \ref{lm:sliding} we quote Lemma 2.1 in
\cite{CLY}. \\
{\bf LEMMA } {\it Let $K: \R^3 \to \R $ be given by
$$
K(z) = r^{-1}\left\{e^{-\nu r} - e^{-\omega r}h(z)\right\}
$$
with $r=|z|$ and $\omega >\nu \geq 0$. Let $h$ satisfy {\rm (i)} $h$ is a $C^4$
function of compact support; {\rm (ii)} $h(z) =1+ar^2 +\mathrm{ O}(r^3)$ near 
$z=0$. Let $h(z)=h(-z)$, so that $K$ has a real Fourier transform.
Then there is a constant, $C_3$ (depending on $h$) such that 
if $\omega - \nu\geq C_3$ then $K$ has a positive Fourier transform and,
moreover, 
$$
\sum_{1\leq i<j\leq N}e_ie_j K(x_i-x_j) \geq \frac{1}{2}(\nu -\omega)N
$$
for all $x_1,\dots x_N \in \R^3$ and all $e_i =\pm1$.     
}

\begin{lm}[Electrostatic
  decoupling of boxes using sliding]\label{lm:sliding}\hfill\\
  There exists a function of the form $\omega(t)=\const t^{-4}$ (we
  assume that $\omega(t)\geq1$ for $t<1/2$) and a constant $\gamma$
  with $1\leq \gamma\leq (1-2t)^{-3}$ such that if we set
  \begin{equation}\label{eq:w} w(x,y)=
    \upchi_\ell(x) Y_{{\omega(t)}/{\ell}}(x-y)\upchi_\ell(y) 
  \end{equation} 
  then the potential energy satisfies 
  \begin{eqnarray*} \lefteqn{U_{pp} + U_{pb} +U_{bb}} &&\\
    &\displaystyle\geq\gamma\sum_{\lambda\in
      \Z^3}\int\limits_{\mu\in[-\frac{1}{2},\frac{1}{2}]^3}\!\!\!\! d\mu\Bigl\{&
    \sum_{1\leq i<j\leq N}w\left(x_i+(\mu+\lambda)\ell,
      x_j+(\mu+\lambda)\ell\right)\\ 
    &&{}-\rho\sum_{j=1}^N  \int_{\Omega}
    w\left(x_j+(\mu+\lambda)\ell ,y+(\mu+\lambda)\ell\right)\,dy \\
    &&{}+\mfr{1}{2}\rho^2\iint_{\Omega\times\Omega}
    w\left(x+(\mu+\lambda)\ell,y+(\mu+\lambda)\ell\right)\,dx\,dy\Bigr\}
    -\frac{\omega(t) N}{2\ell}.  \end{eqnarray*} 
\end{lm} 
\begin{proof} We    
  calculate
  \begin{eqnarray*} \lefteqn{\sum_{\lambda\in
        \Z^3}\int\limits_{\mu\in[-1/2,1/2]^3}\!\!\!\!\!\! d\mu \,
      \gamma\upchi(x+(\mu+\lambda))Y_\omega(x-y)\upchi(y+(\mu+\lambda))}
    \hspace{2cm}&& \\ &=&\int
    \gamma\upchi(x+z)Y_\omega(x-y)\upchi(y+z)\,dz
    =h(x-y)Y_\omega(x-y), 
  \end{eqnarray*} 
  where we have set $h=\gamma\upchi*\upchi$.  Note that $h(0)=1$ and that
  $h$ satisfies all the assumptions in
  Lemma~2.1 in \cite{CLY}.  We then conclude from Lemma~2.1 in
  \cite{CLY} that the Fourier transform of the function
  $F(x)=|x|^{-1}-h(x)Y_{\omega(t)}(x)$ is non-negative, where $\omega$
  is a  function such that $\omega(t)\to\infty$ as $t\to0$.  [The detailed
  bounds from \cite{CLY} show that we may in fact choose
  $\omega(t)=\const t^{-4}$, since $\omega(t)$ has to control the 4th
  derivative of $h$.] Note, moreover, that $\lim_{x\to0}F(x)=\omega(t)$.
  Hence
  $$ \sum_{1\leq i<j\leq N}F(y_i-y_j) -\rho\sum_{j=1}^N
  \int_{\ell^{-1}\Omega}\!\!\!\! F(y_j-y)\,dy
  +\mfr{1}{2}\rho^2\!\!\!\!\iint_{\ell^{-1}\Omega\times\ell^{-1}\Omega}
  \!\!\!\!F(x-y)\,dx\,dy \geq -\frac{N\omega(t)}{2}.  $$

The lemma follows by writing $|x|^{-1} = F(x) +h(y) Y_{\omega(t)}(x)$
and by rescaling from boxes of size 1 to boxes of size $\ell$. 
\end{proof}

As explained above we shall choose the 
parameters $t$ and $\ell$ as functions of $\rho$ at the very end of
the proof. We shall choose them in such a way that 
$t\to0$ and $\rho^{1/4}\ell\to\infty$ as $\rho\to\infty$. 
Moreover, we will have conditions of the form
$$
\rho^{-\tau}(\rho^{1/4}\ell)\to0,\quad \hbox{and}\quad
t^\nu(\rho^{1/4}\ell)\to\infty
$$
as $\rho\to\infty$, where $\tau,\nu$ are universal constants.
 
Consider now the $n$-particle Hamiltonian 
\begin{equation}\label{eq:Hnmu}
  H^{n}_{\mu,\lambda}=-\mfr{1}{2}\sum_{j=1}^n\Delta^{(j)}_{Q_{\mu,\lambda}}
  +\gamma W_{\mu,\lambda},
\end{equation}
where we have introduced the Neumann Laplacian  
$\Delta^{(j)}_{Q_{\mu,\lambda}}$ of the cube
$Q_{\mu,\lambda}=(\mu+\lambda)\ell+
\left[-\frac{1}{2}\ell,\frac{1}{2}\ell\right]^3$ and the potential
\begin{eqnarray*}
  W_{\mu,\lambda}(x_1,\ldots,x_n)&=&
  \sum_{1\leq i<j\leq n}w\left(x_i+(\mu+\lambda)\ell,
    x_j+(\mu+\lambda)\ell\right)\\
  &&-\rho\sum_{j=1}^n  \int_{\Omega} w\left(x_j+(\mu+\lambda)\ell
    ,y+(\mu+\lambda)\ell\right)\,dy \\
  &&+\mfr{1}{2}\rho^2\iint_{\Omega\times\Omega} 
  w\left(x+(\mu+\lambda)\ell,y+(\mu+\lambda)\ell\right)\,dx\,dy.
\end{eqnarray*}

\begin{lm}[Decoupling of boxes]\hfill\\ 
  Let $E^{n}_{\mu,\lambda}$ be the ground state energy of the
  Hamiltonian $H^{n}_{\mu,\lambda}$ given in (\ref{eq:Hnmu})
  considered as a bosonic Hamiltonian.  The ground state energy $E_0$
  of the Hamiltonian $H$ in (\ref{ham}) is then bounded below as
  $$
  E_0\geq\sum_{\lambda\in\Z^3}\int\limits_{\mu\in[-\frac{1}{2},\frac{1}{2}]^3}
  \!\!\!\!\inf_{1\leq n\leq N}E^{n}_{\mu,\lambda}\,d\mu 
  -\frac{\omega(t)N}{2\ell}.
  $$
\end{lm}
\begin{proof}
  If $\Psi(x_1,\ldots,x_N)\in L^2(\R^{3N})$ is a 
  symmetric function. Then 
  $$
  \left(\Psi, H\Psi\right)\geq 
  \sum_{\lambda\in\Z^3}\int\limits_{\mu\in[-\frac{1}{2},\frac{1}{2}]^3}
  \!\!\!\! (\Psi,\widetilde{H}_{\mu,\lambda}\Psi)\,d\mu
  -\frac{\omega(t)N}{2\ell},
  $$
  where 
  \begin{eqnarray*}
    (\Psi,\widetilde{H}_{\mu,\lambda}\Psi)&=&
    \sum_{j=1}^N\int_{x_j\in Q_{\mu,\lambda}}
    |\nabla_j\Psi(x_1,\ldots,x_N)|^2\,dx_1\ldots dx_N\\
    &&{}+\gamma\int W_{\mu,\lambda}(x_1,\ldots,x_N)|\Psi(x_1,\ldots,x_N)|^2
    \,dx_1\ldots dx_N.
  \end{eqnarray*}
  The lemma follows since it is clear that
  $(\Psi,\widetilde{H}_{\mu,\lambda}\Psi)\geq \inf_{1\leq n\leq N}
  E^{n}_{\mu,\lambda}$
\end{proof} 

For given $\mu$
the Hamiltonians $H^n_{\mu,\lambda}$ 
fall in three groups depending on $\lambda$. 
The first kind for which $Q_{\lambda,\mu}\cap \Omega=\emptyset$.
They describe boxes with no background. The optimal energy for these 
boxes are clearly achieved for $n=0$.
The second kind for which $Q_{\lambda,\mu}\subset \Omega$. 
These Hamiltonians are all unitarily equivalent to 
$\gamma H^n_\ell$, where
\begin{eqnarray}
  H^n_\ell&=&\sum_{j=1}^n\left(-\mfr{1}{2}\gamma^{-1}\Delta_{\ell,j}
    -\rho\int w(x_j,y)\,dy \right)\nonumber
  \\&&{}
  +\sum_{1\leq i<j\leq n}w(x_i,x_j)
  +\mfr{1}{2}\rho^2\iint
  w(x,y)\,dx\,dy,\label{eq:Hnell}
\end{eqnarray}
where $-\Delta_\ell$ is the Neumann Laplacian for
the cube $[-\ell/2,\ell/2]^3$.
Finally, there are operators of the third kind for 
which $Q_{\mu,\lambda}$ intersects both $\Omega$ and its complement. 
In this case the particles only see part of the background.
If we artificially add the missing background 
only the last term in the potential $W_{\mu,\lambda}$ increases.
(The first term does not change and the second can only decrease.) 
In fact it will increase by no more than 
$$
\mfr{1}{2}\rho^2\iint w(x,y)\,dx\,dy\leq 
\mfr{1}{2}\rho^2\!\!\!\!\!\!\iint_{x\in[-\ell/2,\ell/2]^3\atop
  y\in[-\ell/2,\ell/2]^3}\!\!\!|x-y|^{-1}
\,dx\,dy\leq \const \rho^2 \ell^5.
$$
Thus the operator $H^n_{\mu,\lambda}$ of the third kind
are bounded below  by an operator which is unitarily equivalent to 
$\gamma H^n_\ell-\const \rho^2 \ell^5$.

We now note that the number of boxes of the third kind 
is bounded above by $\const (L/\ell)^2$. 
The total number of boxes of the second or third kind
is bounded above by $(L+\ell)^3/\ell^3=(1+L/\ell)^3$. 

We have therefore proved the following result. 
\begin{lm}[Reduction to one small box]\label{lm:reducing}\hfill\\ 
  The ground state energy 
  $E_0$ of the Hamiltonian $H$ in (\ref{ham}) is bounded below 
  as
  $$
  E_0\geq(1+L/\ell)^3\gamma \inf_{1\leq n\leq N}\inf\hbox{ \rm Spec }H^n_\ell
  -\const (L/\ell)^2\rho^2 \ell^5
  -\frac{\omega(t)N}{2\ell},
  $$
  where $H_\ell^n$ is the Hamiltonian defined in (\ref{eq:Hnell}).
\end{lm}  
In the rest of the paper we shall study the Hamiltonian
(\ref{eq:Hnell}).

\section{Long and short distance cutoffs in the potential}\label{sec:cutoff}

The potential in the Hamiltonian (\ref{eq:Hnell})
is $w$ given in (\ref{eq:w}). Our aim in this section to 
replace $w$ by a function that has long and short distance  
cutoffs. 

We shall replace the function $w$ by
\begin{equation}\label{eq:wrR}
  w_{r,R}(x,y)=\upchi_\ell(x)V_{r,R}(x-y)\upchi_\ell(y)
\end{equation}
where 
\begin{equation}\label{eq:VrR}
  V_{r,R}(x)=Y_{R^{-1}}(x)-Y_{r^{-1}}(x)=
  \frac{e^{-|x|/R}-e^{-|x|/r}}{|x|}
\end{equation}
Here $0<r\leq R\leq\omega(t)^{-1}\ell$. 
Note that for $x\ll r$ then 
$V_{r,R}(x)\approx r^{-1}-R^{-1}$ 
and for $|x|\gg R$ then  
$V_{r,R}(x)\approx |x|^{-1}e^{-|x|/R}$.

In this section we shall bound the effect of replacing
$w$ by $w_{r,R}$. We shall not fix the cutoffs $r$ and $R$, 
but rather choose them differently at different stages in the later
arguments. 

We first introduce the cutoff $R$ alone, i.e., we bound the 
effect of replacing $w$ by 
$w_R(x,y)=\upchi_\ell(x)V_R(x-y)\upchi_\ell(y)$, 
where $V_R(x)=|x|^{-1}e^{-|x|/R}=Y_{R^{-1}}(x)$.
Thus, since $R\leq \omega(t)^{-1}\ell$, the Fourier transforms satisfy 
$$
\widehat{Y}_{\omega/\ell}(k)-\widehat{V}_R(k)
=4\pi\left(\frac{1}{k^2+(\omega(t)/\ell)^{2}}
  -\frac{1}{k^2+R^{-2}}\right)\geq0.
$$
(We use the convention that 
    $\hat{f}(k)=\int f(x)e^{-ikx}\,dx$.)
Hence 
$w(x,y)-w_R(x,y)=
\upchi_\ell(x)\left(Y_{\omega/\ell}-V_R\right)(x-y)\upchi_\ell(y)$
defines a positive semi-definite kernel.
Note, moreover, that 
$\left(Y_{\omega/\ell}-V_R\right)(0)=
R^{-1}-\omega/\ell\leq R^{-1}$ 
Thus,  
\begin{eqnarray}
  &&\sum_{1\leq i<j\leq n}w(x_i,x_j)
  -\rho\sum_{j=1}^n  \int w(x_j,y)\,dy +\mfr{1}{2}\rho^2\iint
  w(x,y)\,dx\,dy\nonumber\\
  &&-\left(\sum_{1\leq i<j\leq n}w_R(x_i,x_j)
    -\rho\sum_{j=1}^n  \int w_R(x_j,y)\,dy +\mfr{1}{2}\rho^2\iint
    w_R(x,y)\,dx\,dy\right) 
\nonumber\\
  &=&\frac{1}{2}\iint\left[\sum_i^n\delta(x-x_i)-\rho\right]
 (w-w_r)(x,y)
  \left[\sum_i^n\delta(y-x_i)-\rho\right]\,dx\,dy\nonumber\\
  &&-\frac{1}{2}\sum_i^n\upchi_\ell(x_i)^2\left(Y_{\omega/\ell}-V_R\right)(0)
 \geq  -\mfr{1}{2}n \left(Y_{\omega/\ell}-V_R\right)(0) 
 = -\mfr{1}{2}nR^{-1}.\label{eq:R}
\end{eqnarray}

We now bound the effect of replacing $w_R$ by $w_{r,R}$. 
I.e., we are replacing 
$V_R(x)=|x|^{-1}e^{-|x|/R}$ by
$|x|^{-1}\left(e^{-|x|/R}-e^{-|x|/r}\right)$. 
This will lower the repulsive terms and for the attractive term
we get
\begin{eqnarray}
  -\rho\sum_{j=1}^n  \int w_R(x_j,y)\,dy 
  &\geq& -\rho\sum_{j=1}^n  \int w_{r,R}(x_j,y)\,dy\nonumber\\ 
  &&-n\rho\sup_x\int \upchi_\ell(x)\frac{e^{-|x-y|/r}}{|x-y|}
  \upchi_\ell(y)\,dy\nonumber \\
  &\geq&  -\rho\sum_{j=1}^n  \int w_{r,R}(x_j,y)\,dy
  -\const n\rho r^2.\label{eq:r}
\end{eqnarray}
If we combine the bounds (\ref{eq:R}) and (\ref{eq:r})
we have  the following result. 
\begin{lm}[Long and short distance potential cutoffs]
  \label{lm:cutoffs}\hfill\\
  Consider the Hamiltonian
  \begin{eqnarray}
    H^n_{\ell,r,R}&=&\sum_{j=1}^n\left(-\mfr{1}{2}\gamma^{-1}\Delta_{\ell,j}
      -\rho\int w_{r,R}(x_j,y)\,dy\right)+
    \sum_{1\leq i<j\leq n}w_{r,R}(x_i,x_j)
    \nonumber 
    \\&&{}
    +\mfr{1}{2}\rho^2\iint
    w_{r,R}(x,y)\,dx\,dy,\label{eq:HnellrR}
  \end{eqnarray}
  where $w_{r,R}$ is given in (\ref{eq:wrR}) and (\ref{eq:VrR})
  with $0<r\leq R\leq\omega(t)^{-1}\ell$ and  
  $-\Delta_\ell$ as before is the Neumann Laplacian for 
  the cube $[-\ell/2,\ell/2]^3$.
  Then the Hamiltonian $H^n_\ell$ defined in (\ref{eq:Hnell}) 
  obeys the lower bound 
  $$
  H^n_\ell\geq H^n_{\ell,r,R}-\mfr{1}{2}nR^{-1}-\const_1 n\rho r^2.
  $$ 
\end{lm}
A similar argument gives the following result.
\begin{lm} \label{lm:r'R'}With the same notation as above we have for 
$0<r'\leq r\leq R\leq R'\leq\omega(t)^{-1}\ell$ that 
$$
H^n_{\ell,r',R'}\geq H^n_{\ell,r,R}-\mfr{1}{2}nR^{-1}-\const_1 n\rho r^2.
$$ 
\end{lm}
\begin{proof}
Simply note that 
$
V_{r',R'}(x)-V_{r,R}(x)=Y_{R'^{-1}}(x)-Y_{R^{-1}}(x)+
Y_{r^{-1}}(x)-Y_{r'^{-1}}(x)
$  and now use the same arguments as before.
\end{proof}
\begin{cl}[The particle number $n$ cannot be too small]\label{cl:n>rho}\hfill\\
  There exists a constant $C>0$ such that 
  if $\omega(t)^{-1}\rho^{1/3}\ell>C$ then $H^n_\ell\geq
  0$ if $n\leq C\rho\ell^3$. 
\end{cl}
\begin{proof}
  Choose $R=\rho^{-1/3}$ and $r=\frac{1}{2}R$. Then we may assume that 
  $R\leq \omega(t)^{-1}\ell$ 
  since $\omega(t)^{-1}\rho^{1/3}\ell$ is large.
  {F}rom Lemma~\ref{lm:cutoffs} we see immediately that 
  \begin{eqnarray*}
    H^n_\ell&\geq&-\sum_{j=1}^n\rho\int w_{r,R}(x_j,y)\,dy
    +\mfr{1}{2}\rho^2\iint w_{r,R}(x,y)\,dx\,dy-\const n\rho R^2\\
    &\geq& -n\sup_{x}\rho\int w_{r,R}(x,y)\,dy
    +\mfr{1}{2}\rho^2\iint w_{r,R}(x,y)\,dx\,dy-\const n\rho R^2.
  \end{eqnarray*}
  The corollary follows since 
  $\sup_{x}\int w_{r,R}(x,y)\,dy\leq 4\pi R^2$
  and with the given choice of $R$ and $r$ it is easy to see that
  $
  \mfr{1}{2}\iint w_{r,R}(x,y)\,dx\,dy\geq c R^2 \ell^3
  $.
\end{proof}
\section{Bound on the unimportant part of the Hamiltonian}
\label{sec:estimates}

In this section we shall bound the 
Hamiltonian $H^n_{\ell,r,R}$ given in (\ref{eq:HnellrR}). 
We emphasize that we do not necessarily have neutrality in the 
cube, i.e., $n$ and $\rho\ell^3$ may be different. 
We are simply looking for a lower bound to $H^n_{\ell,r,R}$, 
that holds for all $n$. 
The goal is to find a lower bound that will  
allow us to conclude that the optimal $n$, i.e., 
the value for which the energy of the Hamiltonian is smallest, 
is indeed close to the neutral value. 

We shall express the Hamiltonian in second quantized language. 
This is purely for convenience. We stress that we are not 
in any way changing the model by doing this and the treatment 
is entirely rigorous and could have been done without
the use of second quantization. 

Let $u_p$, $\ell p/\pi\in \left(\N\cup\{0\}\right)^3$ be an orthonormal  
basis of   eigenfunctions of the Neumann Laplacian $-\Delta_\ell$
such that $-\Delta_\ell u_p=|p|^2u_p$. I.e., 
$$
u_p(x_1,x_2,x_3)=c_p\ell^{-3/2}
\prod_{j=1}^3
\cos\left(\frac{p_j\pi(x_j+\ell/2)}{\ell}\right),
$$
where the normalization satisfies $c_0=1$ and in general 
$1\leq c_p\leq \sqrt{8}$.
The function $u_0=\ell^{-3/2}$ is the constant eigenfunction with eigenvalue $0$. 
We note that for $p\ne0$ we have 
\begin{equation}\label{eq:gap}
  (u_p,-\Delta_\ell u_p)\geq \pi^2\ell^{-2}.                  
\end{equation}

We now express the Hamiltonian $H^n_{\ell,r,R}$ in terms
of the creation and annihilation operators 
$\an_p=a(u_p)$ and $a^*_p=a(u_p)^*$. 

Define
$$
\hw_{pq,\mu\nu}=\iint w_{r,R}(x,y)u_p(x)u_q(y)u_\mu(x)u_\nu(y)
\,dx\,dy.
$$

We may then express the two-body repulsive potential as 
$$
\sum_{1\leq i<j\leq n}
w_{r,R}(x_i,x_j)=
\mfr{1}{2}\sum_{pq,\mu\nu}\hw_{pq,\mu\nu}
a^*_pa^*_q\an_\nu \an_\mu,
$$
where the right hand side is considered restricted to the 
$n$-particle subspace.
Likewise the Background potential can be written
$$
-\rho\sum_{j=1}^nw_{r,R}(x_j,y)\,dy
=-\rho\ell^3\sum_{pq}\hw_{0p,0q}a^*_p\an_q
$$
and the background-background energy
$$
\mfr{1}{2}\rho^2\iint
w_{r,R}(x,y)\,dx\,dy=\mfr{1}{2}\rho^2\ell^6\hw_{00,00}.
$$
We may therefore write the Hamiltonian
as
\begin{equation}\label{eq:2ndquantization}
  H^n_{\ell,r,R}=
  \mfr{1}{2}\gamma^{-1}\sum_p|p|^2 a^*_p\an_p
  +\mfr{1}{2}\sum_{pq,\mu\nu}\hw_{pq,\mu\nu}
  a^*_pa^*_q\an_\nu \an_\mu
  -\rho\ell^3\sum_{pq}\hw_{0p,0q}a^*_p\an_q
  +\mfr{1}{2}\rho^2\ell^6\hw_{00,00}.
\end{equation}

We also introduce the operators
$\hn_0=a^*_0\an_0$ and $\hn_+=\sum_{p\ne0}$.
These operators represent the number of particles in 
the condensate state created by $a^*_0$ 
and the number of particle {\it not} in the  condensate. 
Note that on the subspace where the total particle number is 
$n$,  both of these operators are non-negative and $\hn_+=n-\hn_0$.

Using the bounds on the long and short distance cutoffs in
Lemma~\ref{lm:cutoffs} we may immediately prove a simple bound on 
the expectation value of $\hn_+$. 

\begin{lm}[Simple bound on the number of excited particles]\label{lm:sn+bound}\hfill\\
  There is a constant $C>0$ such that if 
  $\omega(t)^{-1}\rho^{1/3}\ell>C$
  then for any state such that the expectation 
  $\langle H^n_\ell\rangle\leq 0$, the expectation of the number of
  excited  particles satisfies 
  $\langle\hn_+ \rangle\leq \const n \rho^{-1/6}
  \left(\rho^{1/4}\ell\right)^2$. 
\end{lm}
\begin{proof}
  We simply choose $r=R=\rho^{-1/3}$ in Lemma~\ref{lm:cutoffs}.
  This is allowed since $R\leq\omega(t)^{-1}\ell$ is ensured from the
  assumption that $\omega(t)^{-1}\rho^{1/3}\ell$ is large. We then obtain 
  $$
  H^n_\ell\geq \sum_{j=1}^n-\mfr{1}{2}\gamma^{-1}\Delta_{\ell,j}
  -\mfr{1}{2}nR^{-1}-\const n\rho
  r^2\geq \sum_{j=1}^n-\mfr{1}{2}\gamma^{-1}\Delta_{\ell,j}-\const n\rho^{1/3}.
  $$
  The bound on  $\langle\hn_+ \rangle$
  follows since the bound on the gap (\ref{eq:gap}) implies that 
  $\langle\sum_{j=1}^n-\Delta_{\ell,j}\rangle
  \geq \langle\hn_+\rangle\pi^2\ell^{-2}$.
\end{proof}
Motivated by Foldy's use of the Bogolubov
approximation it is our goal  to reduce the Hamiltonian  $H^n_{\ell,r,R}$
so that it has only what we call quadratic terms, i.e.,
terms which contain precisely two 
$a^\#_p$ with $p\ne0$.
More precisely, we want to be able to ignore all terms containing 
the coefficients
\begin{itemize}
\item $\hw_{00,00}$.
\item $\hw_{p0,q0}=\hw_{0p,0q}$, where $p,q\ne0$. These terms
  are in fact quadratic, but do not appear in the Foldy Hamiltonian.
  We shall prove that they can also be ignored.
\item $\hw_{p0,00}=\hw_{0p,00}=\hw_{00,p0}=\hw_{00,0p}$,
  where  $p\ne0$. 
\item
  $\hw_{pq,\mu0}=\hw_{\mu0,pq}=\hw_{qp,0\mu}=\hw_{0\mu,qp}$,
  where $p,q,\mu\ne0$.
\item $\hw_{pq,\mu\nu}$, where $p,q,\mu,\nu\ne0$. 
  The sum of all these terms form a non-negative contribution to the
  Hamiltonian and can, when proving a lower bound,
  either be ignored or used to control error terms. 
\end{itemize}

We shall consider these cases one at a time.

\begin{lm}[Control of terms with $\hw_{00,00}$]\label{lm:0000}\hfill\\
  The sum of the terms in $H^n_{\ell,r,R}$ containing $\hw_{00,00}$ 
  is equal to 
  $$
  \mfr{1}{2}\hw_{00,00}\left[\left(\hn_0-\rho\ell^3\right)^2-\hn_0\right]
  =\mfr{1}{2}\hw_{00,00}\left[\left(n-\rho\ell^3\right)^2
    +\left(\hn_+\right)^2-2 \left(n-\rho\ell^3\right)\hn_+ -\hn_0\right].
  $$
\end{lm}
\begin{proof}
  The terms containing $\hw_{00,00}$ are
  $$
  \mfr{1}{2}\hw_{00,00}\left(
    a^*_0a^*_0\an_0\an_0-2\rho\ell^3a_0^*\an_0
    +\rho^2\ell^6\right)
  =\mfr{1}{2}\hw_{00,00}\left(a_0^*\an_0-\rho\ell^3\right)^2
  -\mfr{1}{2}\hw_{00,00}a_0^*\an_0
  $$
  using the commutation relation $[\an_p, a^*_q]=\delta_{p,q}$. 
\end{proof}

\begin{lm}[Control of  terms with $\hw_{p0,q0}$]\label{lm:p0q0}\hfill\\ 
  The sum of the terms in $H^n_{\ell,r,R}$ containing $\hw_{p0,q0}$
  or $\hw_{0p,0q}$ with $p,q\ne0$ is bounded below by 
  $$
  -4\pi[\rho-n\ell^{-3}]_+\hn_+R^2
  -4\pi\hn_+^2\ell^{-3}R^2,
  $$
  where $[t]_+=\max\{t,0\}$. 
\end{lm}
\begin{proof}
  The terms containing $\hw_{p0,q0}$ or $\hw_{0p,0q}$ are
  $$
  \sum_{p\ne0\atop q\ne0}\left(\mfr{1}{2}\hw_{p0,q0}
    a^*_pa^*_0\an_0 \an_q +\mfr{1}{2}\hw_{0p,0q}a^*_0a^*_p\an_q \an_0 
    -\rho\ell^3\hw_{0p,0q}a^*_p\an_q\right)
  =(\hn_0-\rho\ell^3)\sum_{p\ne0\atop q\ne0}\hw_{p0,q0}a^*_p\an_q.
  $$
  Note that $\hn_0$ commutes with 
  $\sum\limits_{p\ne0\atop q\ne0}\hw_{p0,q0}a^*_p\an_q$. 

  We have that 
  $$
  \hw_{p0,q0}=\ell^{-3}\int\int w_{r,R}(x,y)\,dy u_p(x)u_q(x)\,dx.
  $$
  Hence
  \begin{eqnarray*}
    \sum\limits_{p\ne0\atop q\ne0}\hw_{p0,q0}a^*_p\an_q
    &=&\ell^{-3}\int\int w_{r,R}(x,y)\,dy
    \left(\sum_{p\ne0}u_p(x)a^*_p\right)\left(\sum_{p\ne0}u_p(x)a^*_p\right)^*\,dx.\\
    &\leq&\ell^{-3}\sup_{x'}\int w_{r,R}(x',y)\,dy
    \int\left(\sum_{p\ne0}u_p(x)a^*_p\right)\left(\sum_{p\ne0}u_p(x)a^*_p\right)^*\,dx.
    \\&=&\ell^{-3}\sup_{x'}\int w_{r,R}(x',y)\,dy\sum_{p\ne0}a^*_p\an_p
    =\ell^{-3}\sup_{x'}\int w_{r,R}(x',y)\,dy\hn_+.
  \end{eqnarray*}
  Since
  $$
  \sup_x\int w_{r,R}(x,y)\,dy\leq \int V_{r,R}(y)\,dy\leq
  4\pi R^2
  $$
  we obtain the operator inequality
  $$
  0\leq\sum_{p\ne0\atop q\ne0}\hw_{p0,q0}a^*_p\an_q
  \leq4\pi\ell^{-3}R^2\hn_+,
  $$
  and the lemma follows.
\end{proof}

Before treating the last two types of terms
we shall need the following result on the structure
of the coefficients $\hw_{pq,\mu\nu}$.

\begin{lm}\label{lm:hw}
  For all $p',q'\in(\pi/\ell)\left(\N\cup\{0\}\right)^3$ 
  and $\alpha\in\N$
  there exists $J^\alpha_{p'q'}\in\R$ with 
  $J^\alpha_{p'q'}=J^\alpha_{q'p'}$ such that 
  for all $p,q,\mu,\nu\in(\pi/\ell)\left(\N\cup\{0\}\right)^3$
  we have 
  \begin{equation}\label{eq:wpostyp}
    \hw_{pq,\mu\nu}=\sum_\alpha J^\alpha_{p\mu} J^\alpha_{q\nu}.
  \end{equation}
  Moreover we have the operator inequalities
  \begin{equation}\label{eq:p00p'}
    0\leq\sum_{p,p'\ne0}\hw_{pp',00}a^*_p\an_{p'}=\sum_{p,p'\ne0}\hw_{p0,0p'}a^*_p\an_{p'}
    \leq4\pi\ell^{-3}R^2\hn_+
  \end{equation}
  and
  $$
  0\leq\sum_{p,p',m\ne0}\hw_{pm,mp'}a^*_p\an_{p'}\leq r^{-1}\hn_+.
  $$
\end{lm}
\begin{proof}
  The operator ${\cal A}$ with integral kernel $w_{r,R}(x,y)$ is a non-negative
  Hilbert-Schmidt
  operator on $L^2(\R^3)$ with norm less than 
  $\sup_k\hat V_{r,R}(k)\leq 4\pi R^2$. 
  Denote the eigenvalues of ${\cal A}$  by
  $\lambda_\alpha$, $\alpha=1,2,\ldots$ and   
  corresponding orthonormal eigenfunctions by
  $\varphi_\alpha$. We may assume 
  that these functions are real.
 The eigenvalues satisfy 
  $0\leq\lambda_\alpha\leq {4\pi}R^2$. 
  We then have 
  $$
  \hw_{pq,\mu\nu}=\sum_\alpha\lambda_\alpha
  \int u_p(x)u_\mu(x)\varphi_\alpha(x)\,dx
  \int u_q(y)u_\nu(y)\varphi_\alpha(y)\,dy.
  $$
  The identity (\ref{eq:wpostyp})  thus follows with 
  $J_{p\mu}^\alpha=\lambda_\alpha^{1/2}\int u_p(x)u_\mu(x)\varphi_\alpha(x)\,dx$.

  If $P$ denotes the projection onto the constant functions we may also
  consider the operator $(I-P){\cal A}(I-P)$. 
  Denote its eigenvalues and eigenfunctions
  by $\lambda'_\alpha$ and $\varphi'_\alpha$. 
  Then again $0\leq\lambda'_\alpha\leq 4\pi R^2$.
  Hence we may write 
  $$
  \hw_{p0,0p'}=\ell^{-3}\sum_\alpha\lambda'_\alpha
  \int u_p(x)\varphi'_\alpha(x)\,dx
  \int u_{p'}(y)\varphi'_\alpha(y)\,dy.
  $$
  Thus, since all $\varphi'_\alpha$ are orthogonal to constants
  we have 
  \begin{eqnarray*}
    \lefteqn{\sum_{p,p'\ne0}\hw_{p0,0p'}a^*_p\an_{p'}}&& \\
    &=&\ell^{-3}\sum_\alpha \lambda'_\alpha
    \left(\sum_{p\ne0}\int u_p(x)\varphi'_\alpha(x)\,dx\, a^*_p\right) 
    \left(\sum_{p\ne0}\int u_p(x)\varphi'_\alpha(x)\,dx\, a^*_p\right)^*\\
    &=&\ell^{-3}\sum_\alpha \lambda'_\alpha a^*\left(\varphi'_\alpha\right)
    a\left(\varphi'_\alpha\right)
  \end{eqnarray*}
  The inequalities (\ref{eq:p00p'}) follow immediately from this. 
  
  The fact that $\sum_{p,p',m\ne0}\hw_{pm,mp'}a^*_p\an_{p'}\geq 0$ follows
  from the representation (\ref{eq:wpostyp}).
  Moreover, since the kernel $w_{R,r}(x,y)$ is a continuous function
  we have that
  $w_{r,R}(x,x)=\sum_{\alpha}\lambda_\alpha\varphi_\alpha(x)^2$
  for almost all $x$ and hence
  $$
  \sum_{m\ne0}\hw_{pm,mp'}=\int u_p(x)u_{p'}(x)w_{r,R}(x,x)\,dx
  -\hw_{p0,0p'}.
  $$
  We therefore have 
  \begin{eqnarray*}
    \lefteqn{\sum_{p,p',m\ne0}\hw_{pm,mp'}a^*_p\an_{p'}\leq\sum_{p,p'\ne0}
      \int u_p(x)u_{p'}(x)W_{r,R}(x,x)\,dx\,a^*_p\an_{p'}}&&\\
    &=&\int w_{r,R}(x,x)\left(\sum_{p\ne0} u_p(x)a^*_p\right)
    \left(\sum_{p\ne0} u_p(x)a^*_p\right)^*\,dx\\
    &\leq&\sup_{x'}w_{r,R}(x',x')\int\left(\sum_{p\ne0} u_p(x)a^*_p\right)
    \left(\sum_{p\ne0} u_p(x)a^*_p\right)^*\,dx
    =\sup_{x'}w_{r,R}(x',x')\hn_+
  \end{eqnarray*}
  and the lemma follows since $\sup_{x'}w_{r,R}(x',x')\leq r^{-1}$.
\end{proof}

\begin{lm}[Control of terms with $\hw_{p0,00}$]\label{lm:p000}\hfill\\
  The sum of the terms in $H^n_{\ell,r,R}$ containing $\hw_{p0,00}$,
  $\hw_{0p,00}$,$\hw_{00,p0}$, or $\hw_{00,0p}$, with $p\ne0$
  is, for all $\varepsilon>0$, bounded below by
  \begin{equation}\label{eq:p000-1}
    -\varepsilon^{-1}{4\pi}\ell^{-3}R^2 \hn_0\hn_+
    -\varepsilon\hw_{00,00}
    (\hn_0+1-\rho\ell^3)^2,
  \end{equation}
   and by
  \begin{eqnarray}
    &&\sum_{p\ne0}\hw_{p0,00}\left((n-\rho\ell^3)a_p^*\an_0
      +a_0^*\an_p(n-\rho\ell^3)\right)\nonumber\\&&{}
    \qquad-\varepsilon^{-1}{4\pi}\ell^{-3}R^2 \hn_0\hn_+
    -\varepsilon\hw_{00,00}
    (\hn_+-1)^2.\label{eq:p000-2}
  \end{eqnarray} 
\end{lm}
\begin{proof}The terms containing $\hw_{p0,00}$,
  $\hw_{0p,00}$,$\hw_{00,p0}$, or $\hw_{00,0p}$ are
  \begin{eqnarray*}
    &&\sum_{p\ne0}\mfr{1}{2}\hw_{p0,00}\left(2a_p^*a^*_0\an_0\an_0
      +2a_0^*a^*_0\an_0\an_p-2\rho\ell^3a_0^*\an_p-2\rho\ell^3a_p^*\an_0\right)\\
    &=&\sum_{p\ne0}\hw_{p0,00}\left((\hn_0-\rho\ell^3)a_p^*\an_0
      +a_0^*\an_p(\hn_0-\rho\ell^3)\right)\\
    &=&\sum_\alpha \sum_{p\ne0}J^\alpha_{p0}J^\alpha_{00}
    \left(a_p^*\an_0(\hn_0+1-\rho\ell^3)
      +(\hn_0+1-\rho\ell^3)a_0^*\an_p\right).
  \end{eqnarray*}
  In the last term we have used 
  the representation (\ref{eq:wpostyp}) and the commutation relation
  $[\hn_0,a_0]=a_0$. 
  For all $\varepsilon>0$ we get that the above expression is bounded
  below by 
  \begin{eqnarray*}
    \lefteqn{\varepsilon^{-1}\sum_\alpha \sum_{p,p'\ne0}
      J^\alpha_{p0}J^\alpha_{p'0}\hn_0a^*_p \an_{p'}
      -\varepsilon\sum_\alpha \left(J^\alpha_{00}\right)^2
      (\hn_0+1-\rho\ell^3)^2}&&\\
    &=& -\varepsilon^{-1}\sum_{p,p'\ne0}\hw_{p0,0p'}\hn_0a^*_p \an_{p'}
    -\varepsilon\hw_{00,00}
    (\hn_0+1-\rho\ell^3)^2.
  \end{eqnarray*}
  The bound (\ref{eq:p000-1})  follows from (\ref{eq:p00p'}).
  
  The second  bound (\ref{eq:p000-2}) follows in the same 
  way if we notice that the terms containing $\hw_{p0,00}$,
  $\hw_{0p,00}$,$\hw_{00,p0}$, or $\hw_{00,0p}$  may be written as 
  \begin{eqnarray*}
    &&\sum_{p\ne0}\hw_{p0,00}\left((n-\rho\ell^3)a_p^*\an_0
      +a_0^*\an_p(n-\rho\ell^3)\right)\\&&{}\qquad
    +\sum_\alpha \sum_{p\ne0}J^\alpha_{p0}J^\alpha_{00}
    \left(a_p^*\an_0(1-\hn_+)
      +(1-\hn_+)a_0^*\an_p\right).
  \end{eqnarray*}
\end{proof}

\begin{lm}[Control of terms with $\hw_{pq,m0}$]\label{lm:pqm0}\hfill\\
  The sum of the terms in $H^n_{\ell,r,R}$ containing $\hw_{pq,m0}$,
  $\hw_{pq,0m}$,$\hw_{p0,qm}$, or $\hw_{0p,qm}$, with $p,q,m\ne0$
  is bounded below by
  $$
  -\varepsilon^{-1}{4\pi} \ell^{-3}R^2\hn_0\hn_+ 
  -\varepsilon\hn_+r^{-1} -\varepsilon\sum_{p,m,p',m'\ne0}
  \hw_{mp',pm'}a^*_ma^*_{p'}\an_{m'}\an_{p},
  $$
  for all $\varepsilon>0$.
\end{lm}
\begin{proof}The terms containing $\hw_{pq,m0}$,
  $\hw_{pq,0m}$,$\hw_{p0,qm}$, or $\hw_{0p,qm}$ are 
  \begin{eqnarray*}
    \lefteqn{\sum_{pqm\ne0}
      \hw_{pqm0}\left(a^*_pa^*_q\an_m\an_0+a^*_0a^*_m\an_q\an_p\right)}&&\\
    &=&
    \sum_\alpha \left(\left(\sum_{q\ne0}J^\alpha_{q0}a^*_q\an_0\right)
      \left(\sum_{pm\ne0}J^\alpha_{pm}a^*_p\an_m\right)
      +\left(\sum_{pm\ne0}J^\alpha_{pm}a^*_p\an_m\right)^*
      \left(\sum_{q\ne0}J^\alpha_{q0}a^*_q\an_0\right)^*
    \right)\\
    &\geq&
    \begin{array}[t]{ll}\displaystyle
      -\sum_\alpha&\left(\displaystyle
        \varepsilon^{-1}\left(\sum_{q\ne0}J^\alpha_{q0}a^*_q\an_0\right)
        \left(\sum_{q\ne0}J^\alpha_{q0}a^*_0\an_q\right)\right.\\ \\
      &\displaystyle\left.
        \quad +\varepsilon\left(\sum_{pm\ne0}J^\alpha_{pm}a^*_m\an_p\right)
        \left(\sum_{pm\ne0}J^\alpha_{pm}a^*_p\an_m\right)
      \right).
    \end{array}
  \end{eqnarray*}
  Using that $J^\alpha_{pm}=J^\alpha_{mp}$ 
  we may write this as 
  \begin{eqnarray*}
    \lefteqn{-\varepsilon^{-1}\sum_{qq'\ne0}\hw_{q0,0q'}a^*_q\an_{q'}\an_0a^*_0
      -\varepsilon\sum_{p,m,p',m'\ne0}
      \hw_{mp',pm'}a^*_m\an_pa^*_{p'}\an_{m'}}&&\\
    &=&-\varepsilon^{-1}\sum_{qq'\ne0}\hw_{q0,0q'}a^*_q\an_{q'}\an_0a^*_0
    -\varepsilon\sum_{p,m,p',m'\ne0}
    \hw_{mp',pm'}a^*_ma^*_{p'}\an_{m'}\an_{p}\\
    &&-\varepsilon\sum_{p,m,m'\ne0}
    \hw_{mp,pm'}a^*_m\an_{m'}.
  \end{eqnarray*}
  The lemma now follows from Lemma~\ref{lm:hw}.
\end{proof}

\section{Analyzing the quadratic  Hamiltonian}\label{sec:quadratic}

In this section we consider the main part of the Hamiltonian. 
This is the ``quadratic'' Hamiltonian considered by Foldy.
It consists of the kinetic energy and all the terms with the 
coefficients $\hw_{pq,00}$, $\hw_{00,pq}$ $\hw_{p0,0q}$, and
$\hw_{0p,q0}$with $p,q\ne0$, i.e.,
\begin{eqnarray}\label{eq:Foldyhamiltonian}
  H_{\rm Foldy}&=&\mfr{1}{2}\gamma^{-1}\sum_p|p|^2
  a^*_p\an_p\nonumber\\ 
  &&{}
  +\mfr{1}{2}\sum_{pq\ne0}\hw_{pq,00}\left(
    a^*_pa^*_0\an_0 \an_q+a^*_0a^*_p\an_q \an_0+
    a^*_pa^*_q\an_0 \an_0+a^*_0a^*_0\an_p \an_q\right)\\
  &=&\mfr{1}{2}\gamma^{-1}\sum_p|p|^2 a^*_p\an_p
  +\sum_{pq\ne0}\hw_{pq,00}\left(
    a^*_p\an_qa^*_0\an_0 +
    \mfr{1}{2}a^*_pa^*_q\an_0 \an_0+\mfr{1}{2}a^*_0a^*_0\an_p
    \an_q\right).\nonumber
\end{eqnarray}
In order to compute all  the bounds we found it necessary to include 
the first term in (\ref{eq:p000-2}) into the ``quadratic''
Hamiltonian.
We therefore define 
\begin{eqnarray}\label{eq:quadratichamiltonian}
  H_{Q}&=&\mfr{1}{2}\gamma^{-1}\sum_p|p|^2 a^*_p\an_p
  +\sum_{p\ne0}\hw_{p0,00}\left((n-\rho\ell^3)a_p^*\an_0
    +a_0^*\an_p(n-\rho\ell^3)\right)\nonumber\\&&{}
  +\sum_{pq\ne0}\hw_{pq,00}\left(
    a^*_p\an_qa^*_0\an_0 +
    \mfr{1}{2}a^*_pa^*_q\an_0 \an_0+\mfr{1}{2}a^*_0a^*_0\an_p \an_q\right).
\end{eqnarray}
Note that $H_{\rm Foldy}=H_Q$ in the neutral case $n=\rho\ell^3$.
Our goal is to give a lower bound on the ground state energy of the
Hamiltonian $H_Q$.

For the sake of convenience we first enlarge the one-particle Hilbert space
$L^2\left([-\ell/2,\ell/2]^3\right)$. 
In fact, instead of considering the symmetric Fock space over 
$L^2\left([-\ell/2,\ell/2]^3\right)$ 
we now consider the symmetric Fock space over  the one-particle
Hilbert space $L^2\left([-\ell/2,\ell/2]^3\right)\oplus\C$.
Note that the larger Fock space of course contains the original 
Fock space as a subspace. On the larger  space we have 
a new pair of creation and annihilation operators that we denote
$\at_0^*$ and $\atn_0$. These operators merely create vectors in
the $\C$ component of $L^2\left([-\ell/2,\ell/2]^3\right)
\oplus\C$, and so 
commute with all other  operators. 

We shall now write
\begin{equation}
  \atn_p=\left\{
    \begin{array}{ll}
      \an_p,&\hbox{if }p\ne0\\
      \atn_0,&\hbox{if }p=0
    \end{array}\right.
  \quad\hbox{and}\quad
  \at^*_p=\left\{
    \begin{array}{ll}
      a^*_p,&\hbox{if }p\ne0\\
      \at^*_0,&\hbox{if }p=0
    \end{array}\right.
\end{equation}
We now define the Hamiltonian
\begin{eqnarray}\label{eq:quadratichamiltoniant}
  \widetilde{H}_{Q}&=&
  \mfr{1}{2}\gamma^{-1}\sum_p|p|^2 \at^*_p\atn_p
  +\sum_{p}\hw_{p0,00}\left((n-\rho\ell^3)\at_p^*\an_0
    +a_0^*\atn_p(n-\rho\ell^3)\right)\nonumber\\&&{}
  +\sum_{pq}\hw_{pq,00}\left(
    \at^*_p\atn_qa^*_0\an_0 +
    \mfr{1}{2}\at^*_p\at^*_q\an_0 \an_0+\mfr{1}{2}a^*_0a^*_0
    \atn_p \atn_q\right),
\end{eqnarray}
where we no longer restrict $p,q$ to be different from $0$. 
Note that for all states on the larger Fock space for which 
$\langle \at^*_0\atn_0\rangle=0$ we have 
$\langle\widetilde{H}_{Q} \rangle=\langle{H}_{Q} \rangle$. 

For any function $\varphi\in L^2\left([-\ell/2,\ell/2]^3\right)$ 
we introduce the creation operator
$$
\at^*(\varphi)=\sum_p(u_p,\varphi)\at^*_p.
$$
Note that the sum includes $p=0$. 
the difference from $a^*(\varphi)$ is given by
$\at^*(\varphi)-a^*(\varphi)= (u_0,\varphi)\left(\at^*_0-a^*_0\right)$.

Then $[\at(\varphi),\at^*(\psi)]=(\varphi,\psi)$. 
We have introduced the ``dummy'' operator $\at^*_0$ 
in order for this relation to hold. One could just as well have 
stayed in the old space, but then the relation above would hold 
only for functions orthogonal to constants. 

For any $k\in\R^3$ denote $\upchi_{\ell,k}(x)=e^{ikx}\upchi_\ell(x)$ 
and define the operators
$$
b_k^*=\at^*(\upchi_{\ell,k})\an_0\quad
\hbox{and}\quad
\bn_k=\at(\upchi_{\ell,k})a_0^*
$$
They satisfy the commutation relations
\begin{equation}
[\bn_k,b^*_{k'}]=a_0^*\an_0\left(\upchi_{\ell,k},\upchi_{\ell,k'}\right)
-\at(\upchi_{\ell,k})\at^*(\upchi_{\ell,k'})
=a_0^*\an_0\widehat{\upchi_\ell^2}(k'-k)
-\at(\upchi_{\ell,k})\at^*(\upchi_{\ell,k'})\label{eq:bcommutation}
\end{equation}
We first consider the kinetic energy part of the Hamiltonian.  We
shall bound it using the double commutator bound in
 Appendix  \ref{app:double}.
First we need a well known comparisson between the Neumann Laplacian
and
the Laplacian in the whole space. 

\begin{lm}[Neumann resolvent is bigger than free resolvent]
  \label{lm:Neumannresolvent}\hfill\\
  Let $P_\ell$ denote the projection in $L^2(\R^3)$ that projects onto 
  $L^2([-\ell/2,\ell/2]^3)$ (identified as a subspace). Then if $-\Delta$ denotes
  the Laplacian on all of $\R^3$ and $-\Delta_\ell$ is the Neumann
  Laplacian on $[-\ell/2,\ell/2]^3$ we have the operator inequality
  $$
  (-\Delta_\ell+a)^{-1}\geq P_\ell(-\Delta+a)^{-1}P_\ell,
  $$
  for all $a>0$.
\end{lm}
\begin{proof}
  It is clear that for all $f\in L^2(\R^3)$
  $$
  \|P_\ell(-\Delta_\ell+a)^{1/2}P_\ell(-\Delta+a)^{-1/2}f\|^2\leq \|f\|^2,
  $$
  and hence 
  $$
  \|(-\Delta+a)^{-1/2}P_\ell(-\Delta_\ell+a)^{1/2}P_\ell f\|^2\leq \|f\|^2.
  $$
  Now simply use this with $f=(-\Delta_\ell+a)^{-1/2}u$.
\end{proof}
\begin{lm}[The kinetic energy bound]\hfill\\
  There exists a constant $C'>0$ such that if
  $C't<1$, where $t$ is the parameter used in the definition of
  $\upchi_\ell$ in Section~\ref{sec:sliding}, we have
  $$
  \left\langle\sum_p|p|^2 \at^*_p\atn_p\right\rangle
  \geq 
  (2\pi)^{-3}(1-C't)^2n^{-1}\int_{\R^3} 
  \frac{|k|^4}{|k|^2+(\ell t^3)^{-2}}
  \langle b_k^*\bn_k\rangle\,dk
  $$
  for all states with $\langle\at^*_0\atn_0\rangle=0$ and 
  particle number equal to $n$, i.e., 
  $\left\langle\sum_p a^*_p\an_p\right\rangle^2
  =\left\langle\left(\sum_p a^*_p\an_p\right)^2\right\rangle=n^2$.
\end{lm}
\begin{proof}Let $s$, with $0<s\leq t$, be a parameter to be chosen
  below. Recall that $t$ is the parameter used in the definition of
  $\upchi_\ell$ in Section~\ref{sec:sliding}.  Then
  since $\upchi_\ell^2+\eta_\ell^2=1$ we have 
  \begin{eqnarray*}
    -\Delta_\ell&\geq&\frac{(-\Delta_\ell)^2}{-\Delta_\ell+((\ell s)^{-2}}
    =\mfr{1}{2}(\upchi_\ell^2+\eta_\ell^2)
    \frac{(-\Delta_\ell)^2}{-\Delta_\ell+(\ell s)^{-2}}
    +\mfr{1}{2}
    \frac{(-\Delta_\ell)^2}{-\Delta_\ell+(\ell s)^{-2}}
    (\upchi_\ell^2+\eta_\ell^2)\\
    &=&\upchi_\ell \frac{(-\Delta_\ell)^2}{-\Delta_\ell+(\ell s)^{-2}}
    \upchi_\ell+
    \eta_\ell \frac{(-\Delta_\ell)^2}{-\Delta_\ell+(\ell s)^{-2}}\eta_\ell
    \\&&{}
    +\left[\left[\frac{(-\Delta_\ell)^2}{-\Delta_\ell+(\ell s)^{-2}},
        \upchi_\ell\right],\upchi_\ell\right]
    +\left[\left[\frac{(-\Delta_\ell)^2}{-\Delta_\ell+(\ell s)^{-2}},
        \eta_\ell\right],\eta_\ell\right]\\
    &\geq&\upchi_\ell \frac{(-\Delta_\ell)^2}{-\Delta_\ell+(\ell s)^{-2}}
    \upchi_\ell+
    \eta_\ell \frac{(-\Delta_\ell)^2}{-\Delta_\ell+(\ell s)^{-2}}\eta_\ell
    \\&&{}
    -\const (\ell t)^{-2}\frac{-\Delta_\ell}{-\Delta_\ell+(\ell s)^{-2}}
    -\const \ell^{-2}s^2t^{-4},
  \end{eqnarray*}
  where the last inequality follows from Lemma~\ref{lm:double} in
  Appendix~\ref{app:double}. 
  We can now repeat this calculation to get
  \begin{eqnarray*}
    -\Delta_\ell&\geq&\upchi_\ell \left(
      \frac{(-\Delta_\ell)^2}{-\Delta_\ell+(\ell s)^{-2}}
      -\const (\ell t)^{-2}\frac{-\Delta_\ell}{-\Delta_\ell+(\ell s)^{-2}}
    \right)\upchi_\ell\\ &&{}
    +\eta_\ell\left( 
      \frac{(-\Delta_\ell)^2}{-\Delta_\ell+(\ell s)^{-2}}
      -\const (\ell t)^{-2}\frac{-\Delta_\ell}{-\Delta_\ell+(\ell s)^{-2}}
    \right)\eta_\ell-\const \ell^{-2}s^2t^{-4}
    \\&&{}
    -\const (\ell t)^{-2}\left(
    \left[\left[\frac{-\Delta_\ell}{-\Delta_\ell+(\ell s)^{-2}}
        ,\upchi_\ell\right],\upchi_\ell\right]
    +
    \left[\left[\frac{-\Delta_\ell}{-\Delta_\ell+(\ell s)^{-2}}
        ,\eta_\ell\right],\eta_\ell\right]\right).
  \end{eqnarray*}
  If we therefore use (\ref{eq:doublecom1}) in 
  Lemma~\ref{lm:double} and recall that $s\leq t$ we arrive at
  \begin{eqnarray*}
    -\Delta_\ell&\geq&\upchi_\ell \left(
      \frac{(-\Delta_\ell)^2}{-\Delta_\ell+(\ell s)^{-2}}
      -\const (\ell t)^{-2}\frac{-\Delta_\ell}{-\Delta_\ell+(\ell s)^{-2}}
    \right)\upchi_\ell\\ &&{}
    +\eta_\ell\left( 
      \frac{(-\Delta_\ell)^2}{-\Delta_\ell+(\ell s)^{-2}}
      -\const (\ell t)^{-2}\frac{-\Delta_\ell}{-\Delta_\ell+(\ell s)^{-2}}
    \right)\eta_\ell-\const \ell^{-2}s^2t^{-4}.
  \end{eqnarray*}
  Note that for $\alpha>0$ we have 
  $$
  \alpha\frac{(-\Delta_\ell)^2}{-\Delta_\ell+(\ell s)^{-2}}
      -\const (\ell t)^{-2}\frac{-\Delta_\ell}{-\Delta_\ell+(\ell
        s)^{-2}}
      \geq -\const\alpha^{-1}s^2t^{-4}\ell^{-2}.
  $$
  Thus if we also assume that $\alpha<1$ we have 
  \begin{eqnarray*}
    -\Delta_\ell&\geq&(1-\alpha)\upchi_\ell 
    \frac{(-\Delta_\ell)^2}{-\Delta_\ell+(\ell s)^{-2}}\upchi_\ell
    -\const\alpha^{-1}s^2t^{-4}\ell^{-2}.
  \end{eqnarray*}
  Thus if $u$ is a normalized function on $L^2(\R^3)$ which is
   orthogonal to
  constants we have according to the bound on the gap (\ref{eq:gap}) that
  for all $0<\delta<1$ 
  \begin{eqnarray*}
    (u,-\Delta_\ell u)&\geq& 
    (1-\delta)(1-\alpha)\left(u,\upchi_\ell 
      \frac{(-\Delta_\ell)^2}{-\Delta_\ell+(\ell s)^{-2}}\upchi_\ell
      u\right)\\
    &&{}
    -\const(1-\delta)\alpha^{-1}s^2t^{-4}\ell^{-2}+\delta\pi^2\ell^{-2}.
  \end{eqnarray*}
  We choose $\alpha=\delta=C'st^{-2}$ for an appropriately large
  constant $C'>0$ and assume that
  $s$ and $t$ are such that $\delta$ is less than $1$. Then
  $$
  \left(u,-\Delta_\ell u\right)\geq (1- C' st^{-2})^2\left(u,\upchi_\ell 
  \frac{(-\Delta_\ell)^2}{-\Delta_\ell+(\ell s)^{-2}}\upchi_\ell u\right).
  $$
  If we now use Lemma~\ref{lm:Neumannresolvent} we may write this as
  $$
  \left(u,-\Delta_\ell u\right)\geq (1-C'st^{-2})^2\left(u,\upchi_\ell \Delta_\ell
  \frac{1}{-\Delta+(\ell s)^{-2}}\Delta_\ell\upchi_\ell u\right)
  =(1-C'st^{-2})^2\left(u,\upchi_\ell 
  \frac{(-\Delta)^2}{-\Delta+(\ell s)^{-2}}\upchi_\ell u\right),
  $$
  where in the last inequality we have used that 
  $\Delta\upchi=\Delta_\ell\upchi$ and 
  $\upchi\Delta=\upchi\Delta_\ell$.

  We now choose $s=t^3$ and  we may then
  write this inequality in second quantized form 
 as  
  $$ 
  \left\langle\sum_{p}|p|^2\at^*_p\atn_p\right\rangle\geq (2\pi)^{-3}(1-C't)^2\int_{\R^3} 
  \frac{|k|^4}{|k|^2+(\ell t^3)^{-2}}
  \left\langle\at^*(\upchi_{\ell,k})
  \at(\upchi_{\ell,k})\right\rangle dk
  $$
  using that $\left\langle\at^*_0\atn_0\right\rangle=0$. 
  Since we consider only states with particle number $n$ the
  inequality still holds if we insert $n^{-1}\an_0 a^*_0$ as in the 
  statement of the lemma.
\end{proof}
With the same notation as in the above lemma we may write
$$
w_{r,R}(x,y)=(2\pi)^{-3}\int\hat{V}_{r,R}(k)
\upchi_{\ell,k}(x)\overline{\upchi_{\ell,k}(y)}\,dk.
$$
The last two sums in the Hamiltonian (\ref{eq:quadratichamiltoniant}) 
can  therefore be written as
$$
\begin{array}{rl}\displaystyle
  (2\pi\ell)^{-3}\int\hat{V}_{r,R}(k)\Bigl[&\displaystyle (n-\rho\ell^3)
  \ell^{-3/2}\left(\widehat{\upchi}_\ell(k)b^*_k+
    \overline{\widehat{\upchi}_\ell(k)}\bn_k\right)\\ 
  &\displaystyle{}+\mfr{1}{2}\left(b^*_k\bn_k+b^*_{-k}b_{-k}+b^*_{k}b^*_{-k}
    +\bn_k\bn_{-k}\right)\Bigr]\,dk-\sum_{pq}\hw_{pq,00}
    \at^*_p\atn_q.
\end{array}
$$
Note that it is important here that the potential $w_{r,R}$ 
contains the localization function
$\upchi_\ell$. 

Thus, since $\hat{V}_{r,R}(k)=\hat{V}_{r,R}(-k)$
and $\overline{\widehat{\upchi}_\ell(k)}={\widehat{\upchi}_\ell(-k)}$
we have for states with $\left\langle\at^*_0\atn_0\right\rangle=0$
that 
\begin{equation}\label{eq:hQ}
  \left\langle\widetilde{H}_Q\right\rangle 
  \geq \int_{\R^3}\left\langle h_Q(k)\right\rangle\,dk
    -\sum_{pq}\hw_{pq,00}\left\langle\at^*_p\atn_q\right\rangle,
\end{equation}
where
\begin{eqnarray}\label{eq:hQdef}
  h_Q(k)&=&\frac{(1-C't)^2}{4(2\pi)^{3}\gamma n}
  \frac{|k|^4}{|k|^2+(\ell t^3)^{-2}}
  \left(b_k^*\bn_k+b_{-k}^*\bn_{-k}\right)\\ \nonumber \\
  &&{}\nonumber
  \begin{array}{rl}\displaystyle
    +\frac{\hat{V}_{r,R}(k)}{2(2\pi\ell)^3}\Bigl[&\!\!\!\!\displaystyle (n-\rho\ell^3)
    \ell^{-3/2}\left(\widehat{\upchi}_\ell(k)(b^*_k+\bn_{-k})+
      \overline{\widehat{\upchi}_\ell(k)}(\bn_k+b^*_{-k})\right)\\ \\
    &\displaystyle{}+\left(b^*_k\bn_k+b^*_{-k}b_{-k}+b^*_{k}b^*_{-k}
      +\bn_k\bn_{-k}\right)\Bigr].
  \end{array}
\end{eqnarray}
\begin{thm}[Simple case of Bogolubov's method]\label{thm:bogolubov}\hfill\\
For arbitrary constants $\cA\geq\cB>0$ and $\kappa\in\C$ we have 
the inequality 
\begin{eqnarray*}
\cA(b^*_k\bn_k+b^*_{-k}\bn_{-k})+\cB(b^*_kb^*_{-k}+\bn_k\bn_{-k})+
\kappa(b^*_k+\bn_{-k})+\overline{\kappa}(\bn_k+b^*_{-k})\\ 
\geq-\mfr{1}{2}(\cA-\sqrt{\cA^2-\cB^2})
([\bn_{k},b^*_{k}]+[\bn_{-k},b^*_{-k}])-\frac{2|\kappa|^2}{\cA+\cB}.\hspace{1cm}
\end{eqnarray*}
\end{thm}
\begin{proof}
We may complete the square
\begin{eqnarray*}
\lefteqn{\cA(b^*_k\bn_k+b^*_{-k}\bn_{-k})+\cB(b^*_kb^*_{-k}+\bn_k\bn_{-k})+
\kappa(b^*_k+\bn_{-k})+\overline{\kappa}(\bn_k+b^*_{-k})}&&\\
&=&D(b^*_k+\alpha\bn_{-k}+a)(\bn_k+\alpha b^*_{-k}+\overline{a})
+D(b^*_{-k}+\alpha\bn_{k}+\overline{a})(\bn_{-k}+\alpha b^*_{-k}+a)\\
&&-D\alpha^2([\bn_{k},b^*_{k}]+[\bn_{-k},b^*_{-k}])-2D|a|^2,
\end{eqnarray*}
if
$$
D(1+\alpha^2)=\cA,\quad2D\alpha=\cB,\quad aD(1+\alpha)=\kappa.
$$
We choose the solution 
$
\alpha={\cA}/{\cB}-\sqrt{{\cA^2}/{\cB^2}-1}.
$
Hence
$$
D\alpha^2=\cB\alpha/2=\mfr{1}{2}(\cA-\sqrt{\cA^2-\cB^2}),\quad
D|a|^2=\frac{|\kappa|^2}{D(1+\alpha^2+2\alpha)}=
\frac{|\kappa|^2}{\cA+\cB}.
$$
\end{proof}
Usually when applying Bogolubov's method the commutator
$[\bn_{k},b^*_{k}]$ is a positive constant. In this case the lower
bound in the theorem is actually the bottom of the spectrum of the
operator. If moreover, $\cA>\cB$ the bottom is actually an
eigenvalue. In our case the commutator $[\bn_{k},b^*_{k}]$ 
is not a constant, but according to (\ref{eq:bcommutation}) we have
\begin{equation}\label{eq:bcomk=k'}
[\bn_{k},b^*_{k}]\leq \int\upchi_\ell(x)^2\,dxa^*_0a_0\leq \ell^3a^*_0a_0.
\end{equation}
{F}rom this and the above theorem we easily conclude the following
bound. 
\begin{lm}[Lower bound on quadratic Hamiltonian]\hfill\label{lm:lbqh}\\
On the subspace with $n$ particles we have
\begin{eqnarray*}
H_Q&\geq& -In^{5/4}\ell^{-3/4}
-\mfr{1}{2}\left(n-\rho\ell^3\right)^2\hw_{00,00}
-4\pi n^{5/4}\ell^{-3/4}(n\ell)^{-1/4}
\end{eqnarray*}
where 
$I=\mfr{1}{2}(2\pi)^{-3}\int_{\R^3} 
f(k)-(f(k)^2-g(k)^2)^{1/2}\,dk$ with
$$
g(k)=4\pi\frac{1}{k^2+(n^{1/4}\ell^{-3/4}R)^{-2}}
-4\pi\frac{1}{k^2+(n^{1/4}\ell^{-3/4}r)^{-2}}
$$
and
$$
f(k)=g(k)+\mfr{1}{2}\gamma^{-1}{(1-C't)^2}
\frac{|k|^4}{|k|^2+(n^{1/4}\ell^{1/4} t^3)^{-2}}.
$$
\end{lm}
\begin{proof}
We consider a state with $\langle\at^*_0\atn_0\rangle=0$.
Then $\langle H_Q\rangle=\langle\widetilde{H}_Q\rangle$.
We shall use (\ref{eq:hQ}). Note first that 
$$
\left\langle\sum_{pq}\hw_{pq,00}\at^*_p\atn_q\right\rangle=
\left\langle\sum_{p,q\ne0}\hw_{p0,0q}a^*_p\an_q\right\rangle
\leq 4\pi\ell^{-3}R^2 \hn_+\leq 4\pi\ell^{-1} n
$$
by (\ref{eq:p00p'}) and the fact that $R\leq \ell$.
We may of course rewrite $\ell^{-1}
n=n^{5/4}\ell^{-3/4}(n\ell)^{-1/4}$.

By Theorem~\ref{thm:bogolubov}, (\ref{eq:hQdef}) and (\ref{eq:bcomk=k'}) we have
$$
h_Q(k)\geq -(\cA_k-\sqrt{\cA_k^2-\cB_k^2})
n\ell^3-
\frac{\hat{V}_{r,R}(k)^2(n-\rho\ell^3)^2}{2(2\pi)^6\ell^9(\cA_k+\cB_k)}
    \left|\widehat{\upchi}_\ell(k)\right|^2,
$$
where
$$
\cB_k=\frac{\hat{V}_{r,R}(k)}{2(2\pi\ell)^3},\quad
\cA_k=\cB_k+\frac{(1-C't)^2}{4(2\pi)^{3}\gamma n}
\frac{|k|^4}{|k|^2+(\ell t^3)^{-2}}.
$$
Since $\cA_k>\cB_k$ we have that  
$$
h_Q(k)\geq -(\cA_k-\sqrt{\cA_k^2-\cB_k^2})
n\ell^3-
\frac{\hat{V}_{r,R}(k)(n-\rho\ell^3)^2}{2(2\pi)^3\ell^6}
\left|\widehat{\upchi}_\ell(k)\right|^2.
$$
Note that 
\begin{eqnarray*}
\lefteqn{\int\frac{\hat{V}_{r,R}(k)(n-\rho\ell^3)^2}{2(2\pi)^3\ell^6}
\left|\widehat{\upchi}_\ell(k)\right|^2\,dk}&&\\
&=&\mfr{1}{2}\left(\frac{n}{\ell^3}-\rho\right)^2
\iint\upchi_\ell(x)V_{r,R}(x-y)\upchi_\ell(y)\,dx\,dy
=\mfr{1}{2}\left(n-\rho\ell^3\right)^2\hw_{00,00}.
\end{eqnarray*}
The lemma now follows from (\ref{eq:hQ}) by a simple change of
variables in the $k$ integral.
\end{proof}
As a consequence we get the following bound for the 
Foldy Hamiltonian.
\begin{cl}[Lower bound on the Foldy Hamiltonian]\label{cl:lbfh}\hfill\\
The Foldy Hamiltonian in (\ref{eq:Foldyhamiltonian})  satisfies
\begin{equation}\label{eq:lbfh1}
H_{\rm Foldy}\geq -In^{5/4}\ell^{-3/4}
-4\pi n^{5/4}\ell^{-3/4}(n\ell)^{-1/4}.
\end{equation}
There is cosntant $C>0$ such that if 
$\rho^{1/4}R>C$, $\rho^{1/4}\ell t^3>C$, and 
$t<C^{-1}$ then the  
Foldy Hamiltonian  satisfies the bound
\begin{eqnarray}\label{eq:lbfh2}
H_{\rm Foldy}&\geq& \mfr{1}{4}\sum_p|p|^2a^*_p\an_p-\const n^{5/4}\ell^{-3/4}.
\end{eqnarray}
\end{cl}
\begin{proof}
Lemma~\ref{lm:lbqh} holds for all $\rho$ hence also if we had replaced
$\rho$ by $n/\ell^3$ in this case we get (\ref{eq:lbfh1}).

The integral $I$ satisfies the bound
$$
 I\leq\mfr{1}{2}(2\pi)^{-3}\int_{\R^3}\max\left\{g(k),\mfr{1}{2}g(k)^2(f(k)-g(k))^{-1}\right\}
 \,dk.
$$
By Corollary~\ref{cl:n>rho} we may assume that
$n\geq c\rho\ell^3$. Hence $I$ is bounded by a constant 
as long as $\rho^{1/4}R$ and $\rho^{1/4}\ell t^3$ are sufficiently large and 
$t$ is sufficiently small (which also ensures that $\gamma$ is close
to $1$). Note that we do not have to make any
assumptions on $r$. Moreover, if this is true we also have that 
$n\ell\geq c\rho\ell^4$ is large and hence $(n\ell)^{-1}$ is
small. This would give the bound in the corollary except for the
first positive term. 
The above argument, however, also holds (with different constants) 
if we replace the kinetic energy in the
Foldy Hamiltonian by 
$\mfr{1}{2}\left(\gamma^{-1}-\mfr{1}{2}\right)\sum_p|p|^2a^*_p\an_p$ (assuming that 
$\gamma<2$). This  proves the corollary. 
\end{proof}

Note that if 
\begin{equation}\label{eq:paramlimits}
  n^{1/4}\ell^{-3/4}R\to\infty,\
  n^{1/4}\ell^{-3/4}r\to0,\
  n^{1/4}\ell^{1/4}t^3\to\infty,\hbox{ and }  t\to0
\end{equation}
it follows by dominated convergence that $I$ converges to 
\begin{eqnarray*}
\lefteqn{\mfr{1}{2}(2\pi)^{-3}\int_{\R^3} 
4\pi|k|^{-2}+\mfr{1}{2}|k|^2-
\left((4\pi|k|^{-2}+\mfr{1}{2}|k|^2)^2-(4\pi|k|^{-2})^2\right)^{1/2}
\,dk}&&\\
&=&(2/\pi)^{3/4}\int_0^\infty
1+x^4-
x^2\left(x^4+2\right)^{1/2}
\,dx=-\left(\frac{4\pi}{3}\right)^{1/4}A,
\end{eqnarray*}
where $A$ was given in (\ref{int}). 
Thus if we can show that $n\sim\rho\ell^3$ we see that the 
term  $-In^{5/4}\ell^{-3/4}\sim-I\rho^{1/4} n$ agrees with Foldy's
calculation (\ref{gse}) for the little box of size $\ell$. 

Our task is now to show that indeed $n\sim\rho\ell^3$, i.e.,
that we have approximate neutrality in each little box and that 
the term above containing the integral $I$ is indeed 
the leading term.  

\section{Simple bounds on $n$ and $\hn_+$.}\label{sec:nn+}
The Lemmas 
\ref{lm:cutoffs},\ref{lm:0000},\ref{lm:p0q0},\ref{lm:p000}, and
\ref{lm:pqm0}
together with Lemma~\ref{lm:lbqh} or Corollary~\ref{cl:lbfh} 
control all terms in the Hamiltonian $H^n_\ell$
except the positive term 
$$
\mfr{1}{2}\sum_{p,m,p',m'\ne0}\hw_{mp',pm'}a^*_ma^*_{p'}\an_{m'}\an_{p}.
$$
If we use (\ref{eq:lbfh2}) in Corollary~\ref{cl:lbfh} together with 
the other bounds
we obtain the following bound
if $\rho^{1/4}R$ and $\rho^{1/4}\ell t^3$ are sufficiently large and 
$t$ is sufficiently small

\begin{eqnarray*}
  H^n_\ell&\geq& \mfr{1}{4}\sum_p|p|^2a^*_p\an_p-\const
  n^{5/4}\ell^{-3/4}
  -\mfr{1}{2}nR^{-1}-\const n\rho r^2 \\&&
  +\mfr{1}{2}\hw_{00,00}\left[\left(\hn_0-\rho\ell^3\right)^2
    -\hn_0\right]
  \\&&
  -4\pi[\rho-n\ell^{-3}]_+\hn_+R^2
  -4\pi\hn_+^2\ell^{-3}R^2
  \\&&
  -\varepsilon^{-1}{8\pi}\ell^{-3}R^2 \hn_0\hn_+
  -\varepsilon\hw_{00,00}
  (\hn_0+1-\rho\ell^3)^2
  \\&& 
  -\varepsilon\hn_+r^{-1} 
  +(\mfr{1}{2}-\varepsilon)\sum_{p,m,p',m'\ne0}
  \hw_{mp',pm'}a^*_ma^*_{p'}\an_{m'}\an_{p}.
\end{eqnarray*}
The assumptions on $\rho^{1/4}R$, $\rho^{1/4}\ell t^3$, and $t$ are
needed in order to bound the integral $I$ above by a constant. 
If we choose $\varepsilon=1/4$, use $\hw_{00,00}\leq4\pi R^2\ell^{-3}$ 
and ignore the last positive term in the bound above we arrive at
\begin{eqnarray}
  H^n_\ell&\geq& \mfr{1}{4}\sum_p|p|^2a^*_p\an_p-\const
  n^{5/4}\ell^{-3/4}
  -\mfr{1}{2}nR^{-1}-\const n\rho r^2 
  +\mfr{1}{4}\hw_{00,00}\left(\hn_0-\rho\ell^3\right)^2
  \nonumber\\&&
  -4\pi[\rho-n\ell^{-3}]_+\hn_+R^2
  -4\pi\hn_+^2\ell^{-3}R^2\nonumber
  \\&&
  -{32\pi}\ell^{-3}R^2 \hn_0\hn_+
  -4\pi R^2\ell^{-3}\left(\hn_0-\mfr{1}{2}\rho\ell^3+\mfr{1}{4}\right)
  -\mfr{1}{4}\hn_+r^{-1} 
   \nonumber\\
  &\geq& \mfr{1}{4}\sum_p|p|^2a^*_p\an_p-\const
  n^{5/4}\ell^{-3/4}
  -\mfr{1}{2}nR^{-1}-\const n\rho r^2 
  +\mfr{1}{4}\hw_{00,00}\left(\hn_0-\rho\ell^3\right)^2
   \nonumber\\&&
  -48\pi\ell^{-3} R^2 n\hn_+
  -4\pi R^2\ell^{-3}\left(\hn_0+\mfr{1}{4}\right)
  -\mfr{1}{4}\hn_+r^{-1},\label{eq:l1}
\end{eqnarray}
where in the last inequality we have used that $\rho\ell^3\leq 2n$,
$\hn_0\leq n$ and $\hn_+\leq n$.

\begin{lm}[Simple bound on $n$]\label{lm:sbn}\hfill\\
Let $\omega(t)$ be the function described in 
Lemma~\ref{lm:sliding}. There is a constant $\const>0$ 
such that if $(\rho^{1/4}\ell)t^3>\const$  and 
$(\rho^{1/4}\ell)\rho^{-1/12}$, $t$, and $\omega(t)(\rho^{1/4}\ell)^{-1}$
are smaller than $\const^{-1}$ then for any state 
with $\langle H^n_\ell\rangle \leq0$ we have
$\const^{-1}\rho\ell^3\leq n\leq\const\rho\ell^3$.
\end{lm}
\begin{proof}
The lower bound follows from Corollary~\ref{cl:n>rho}. 
To prove the upper bound on $n$ 
we choose $R=\omega(t)^{-1}\ell$ (the maximally allowed value) and 
$r=b\omega(t)^{-1}\ell$, where we shall choose $b$ sufficiently small, 
in particular $b<1/2$. We then have that $\rho^{1/4}
R=\omega(t)^{-1}\rho^{1/4}\ell$ is large.
Moreover $\hw_{00,00}\geq \const R^2\ell^{-3}=\const \omega(t)^{-2}\ell^{-1}$ 
for some constant $\const>0$ and 
we get from (\ref{eq:l1}) and Lemma~\ref{lm:sn+bound} that
\begin{eqnarray*}
  \langle H^n_\ell\rangle &\geq& \ell^{-1}[-\const
  n^{5/4}\ell^{1/4}
  -\mfr{1}{2}n\omega(t)-\const b^2\omega(t)^{-2} n^2 
  +\const \omega(t)^{-2}\left(\langle\hn_0\rangle -\rho\ell^3\right)^2
  \\&&
  -48\pi  \omega(t)^{-2} \rho^{-1/6}(\ell\rho^{1/4})^2n^2
  -4\pi \omega(t)^{-2}\left(n+\mfr{1}{4}\right)
  -\mfr{1}{4}nb^{-1}\omega(t)],
\end{eqnarray*}
where we have again used that $c\rho\ell^3\leq n$,
$\hn_0\leq n$ and $\hn_+\leq n$.
Note that 
$$
n^{5/4}\ell^{1/4}\leq\const\omega(t)^{-2}n^2(\rho^{1/4}\ell)^{-2}\rho^{-1/4}
\omega(t)^{2}
$$
and  $n\omega(t)\leq \const\omega(t)^{-2}n^2
\rho^{-1}\omega(t)^3$
{F}rom Lemma~\ref{lm:sn+bound} we know that 
$\langle\hn_0\rangle\geq n(1-\const\rho^{-1/6}(\ell\rho^{1/4})^2)$. 
By choosing $b$ small enough we see immediately that $n\leq C\rho\ell^3$.
\end{proof}
Using this result as an input in (\ref{eq:l1}) we can get a better
bound
on $n$ than above and a better bound on $\langle \hn_+\rangle$ than
given in Lemma~\ref{lm:sn+bound}. In particular, the next lemma in fact implies that 
we have near neutrality, i.e., that $n$ is nearly $\rho\ell^3$. 
\begin{lm}[Improved bounds on $n$ and $\langle\hn_+\rangle$]\label{lm:neutralcondensation}
\hfill\\
There exists a constant $\const>0$ such that if 
$(\rho^{1/4}\ell) t^3>C$  and 
$(\rho^{1/4}\ell)\rho^{-1/12} $, $t$, and $\omega(t)(\rho^{1/4}\ell)^{-1}$
are smaller than $\const^{-1}$
then for any state with $\langle H^n_\ell\rangle \leq0$ we have
$\langle\sum_p|p|^2a^*_p\an_p\rangle\leq
\const \rho^{5/4}\ell^3(\rho^{1/4}\ell)$ and 
$$
\langle\hn_+\rangle\leq \const n\rho^{-1/4}(\rho^{1/4}\ell)^3
\quad\hbox{and}\quad
\left(\frac{n-\rho\ell^3}{\rho\ell^3}\right)^2\leq\const  \rho^{-1/4}(\rho^{1/4}\ell)^3.
$$ 
For any other state with $\langle H^n_{\ell,r',R'}\rangle'\leq0$ we
have the same bound on $\langle \hn_+\rangle'$ if
$r'\leq\rho^{-3/8}(\rho^{1/4}\ell)^{1/2}$ and 
$R'\geq a(\rho^{1/4}\ell)^{-2}\ell$ where $a>0$ is an appropriate constant. 
\end{lm}
\begin{proof}
Inserting the bound $n\leq C\rho\ell^3$ into (\ref{eq:l1}) gives
\begin{eqnarray*}
  H^n_\ell
  &\geq& \mfr{1}{4}\sum_p|p|^2a^*_p\an_p-\const
  \rho^{5/4}\ell^{3}
  -\mfr{1}{2}\rho\ell^3R^{-1}-\const \rho^2\ell^3 r^2 
  +\mfr{1}{4}\hw_{00,00}\left(\hn_0-\rho\ell^3\right)^2
  \\&&
  -\const R^2 \rho\hn_+
  -\const R^2\left(\rho+\mfr{1}{4}\ell^{-3}\right)
  -\mfr{1}{4}\hn_+r^{-1}.
\end{eqnarray*}
We now choose $r=\rho^{-3/8}(\rho^{1/4}\ell)^{1/2}$ and $R=a(\rho^{1/4}\ell)^{-2}\ell$,
where we shall choose $a$ below, independently of $\rho$, $\rho^{1/4}\ell$,  and $t$. 
Note that since $\omega(t)(\rho^{1/4}\ell)^{-2}$ is small we may
assume that $R\leq\omega(t)^{-1}\ell$ as required and 
since $(\rho^{1/4}\ell)\rho^{-1/12}$ is small we may assume that
$r\leq R$. 
Moreover
$r^{-1}=\rho^{-1/8}(\rho^{1/4}\ell)^{3/2}\ell^{-2}$ and 
$R^2\rho=a^2(\rho^{1/4}\ell)^{-4}\ell^2\rho=a^2\ell^{-2}$. 
Hence, since $\sum_p|p|^2a^*_p\an_p\geq \pi^2\ell^{-2}\hn_+$ (see
\ref{eq:gap}), we have
\begin{eqnarray*}
  H^n_\ell
  &\geq&
 \mfr{1}{8}\sum_p|p|^2a^*_p\an_p+
 \left(\mfr{\pi^2}{8}
   -a^2-\mfr{1}{4}\rho^{-1/8}(\rho^{1/4}\ell)^{3/2}\right)\ell^{-2}\hn_+
  +\mfr{1}{4}\hw_{00,00}\left(\hn_0-\rho\ell^3\right)^2
   \\&&-(\mfr{1}{2a}+\const)\rho^{5/4}\ell^3(\rho^{1/4}\ell)
  -\const a^2\rho^{5/4}\ell^{3}(\rho^{1/4}\ell)^{-5}(1+(\rho^{1/4}\ell)^{-3}\rho^{-1/4}).
\end{eqnarray*}
By choosing $a$ appropriately (independently of $\rho$,
$\rho^{1/4}\ell$, and $t$)
we immediately get the  
bound on $\langle\sum_p|p|^2a^*_p\an_p\rangle$ and the bound 
$\ell^{-2}\langle\hn_+\rangle\leq\const \rho^{5/4}\ell^3(\rho^{1/4}\ell)$, which implies
the stated bound on $\langle\hn_+\rangle$. 
The bound on $(n-\rho\ell^3)^2(\rho\ell^3)^{-2}$ follows since 
we also have
$\hw_{00,00}\langle\left(\hn_0-\rho\ell^3\right)^2\rangle\leq
\const \rho^{5/4}\ell^3(\rho^{1/4}\ell)$ and
\begin{eqnarray*}
\hw_{00,00}\langle\left(\hn_0-\rho\ell^3\right)^2\rangle&\geq&
\const R^2\ell^{-3}\left(\langle\hn_0\rangle-\rho\ell^3\right)^2\\
&\geq &\const a^2(\rho^{1/4}\ell)^{-4}\ell^2
\left(n-\rho\ell^3-n\const\rho^{-1/4}(\ell\rho^{1/4})^3\right)^2,
\end{eqnarray*}
where we have used the bound on $\langle\hn_+\rangle$ which we have 
just proved. 

The case when $\langle H^n_{\ell,r',R'}\rangle'\leq0$ follows in the
same way because we may everywhere replace $H^n_\ell$ by
$H^n_{\ell,r',R'}$
and use Lemma~\ref{lm:r'R'} instead of Lemma~\ref{lm:cutoffs}.
Note that in this case we already know the bound on $n$ since we still 
assume the existence of the state such that 
$\langle H^n_{\ell}\rangle\leq0$.
\end{proof}

\section{Localization of $\hn_+$.}\label{sec:local}
Note that Lemma~\ref{lm:neutralcondensation} may be interpreted as saying that we have 
neutrality and condensation, in the sense that $\langle \hn_+\rangle$
is a small fraction of $n$, in each little box. 
Although this bound on $\langle \hn_+\rangle$ is sufficient for our 
purposes we still need to know that $\langle \hn_+^2\rangle\sim
\langle \hn_+\rangle^2$. 
We shall however not prove this for a general state with negative
energy. Instead we shall show that we may
change the ground state, without changing its energy expectation
significantly, in such a way that the possible $\hn_+$ values are
bounded by $\const n\rho^{-1/4}(\rho^{1/4}\ell)^3$. 
To do this we shall use the method of localizing large matrices
in Lemma~\ref{local} of Appendix~\ref{app:local}.

We begin with any normalized $n$-particle wavefunction 
$\Psi$ of the operator
$H^n_\ell$. Since $\Psi$ is an $n$-particle wave function
we may write $\Psi=\sum_{m=0}^n c_m\Psi_m$, where for all $m=1,2,\ldots,n$, $\Psi_m$, 
is a normalized eigenfunctions of $\hn_+$ with eigenvalue $m$.  
We may now consider the $(n+1)\times(n+1)$ Hermitean matrix ${\cal A}$
with matrix elements ${\cal A}_{mm'}=\left(\Psi_m,H^n_{\ell,r,R}\psi_m'\right)$.

We shall use Lemma~\ref{local} for this matrix and the vector 
$\psi=(c_0,\ldots,c_n)$. We shall choose $M$ in  Lemma~\ref{local} 
to be of the order of the upper bound
on $\langle\hn_+\rangle$ derived in
Lemma~\ref{lm:neutralcondensation},
e.g., $M$ is the integer part of $n\rho^{-1/4}(\rho^{1/4}\ell)^3$.
Recall that with the assumption in Lemma~\ref{lm:neutralcondensation}
we have $M\gg1$.  
With the notation in
Lemma~\ref{local}
we have $\lambda=(\psi,{\cal A}\psi)=(\Psi,H^n_{\ell,r,R}\Psi)$. 
Note also that because of the structure of $H^n_{\ell,r,R}$ 
we have, again with  the notation in
Lemma~\ref{local}, that $d_k=0$ if $k>3$.
We conclude from Lemma~\ref{local} that there exists a normalized 
wavefunction $\widetilde{\Psi}$ with the property that the corresponding
$\hn_+$ values belong to an interval of length $M$ and such that 
$$
\left(\Psi,H^n_{\ell,r,R}\Psi\right)
\geq \left(\widetilde{\Psi},H^n_{\ell,r,R}\widetilde{\Psi}\right)
-\const M^{-2}(|d_1|+|d_2|).
$$
We shall discuss $d_1,d_2$, which depend on $\Psi$, in detail below,
but first we give the result
on the localization of $\hn_+$ that we shall use. 
\begin{lm}[Localization of $\hn_+$]\label{lm:n+localization}\hfill\\
There is a constant $\const>0$ with the following property. 
If
$(\rho^{1/4}\ell) t^3>\const$ and 
$(\rho^{1/4}\ell)\rho^{-1/12} $, $t$, and
$\omega(t)(\rho^{1/4}\ell)^{-1}$ are less than $\const^{-1}$ and
$r\leq\rho^{3/8}(\rho^{1/4}\ell)^{1/2}$,
$R\geq \const(\rho^{1/4}\ell)^{-2}\ell$ ,
and $\Psi$ is a normalized wavefunction such that  
\begin{equation}\label{eq:Psiassumption}
  \left(\Psi,H^n_{\ell,r,R}\Psi\right)\leq0
\quad\hbox{and}\quad 
\left(\Psi,H^n_{\ell,r,R}\Psi\right)\leq-
\const (n\rho^{-1/4}(\rho^{1/4}\ell)^3)^{-2}(|d_1|+|d_2|)
\end{equation}
then there exists a 
normalized wave function $\widetilde{\Psi}$, which
is a linear combination of eigenfunctions of $\hn_+$ with eigenvalues
less than $\const n\rho^{-1/4}(\rho^{1/4}\ell)^3$ only, such that
\begin{equation}\label{eq:n+localization}
\left(\Psi,H^n_{\ell,r,R}\Psi\right)
\geq \left(\widetilde{\Psi},H^n_{\ell,r,R}\widetilde{\Psi}\right)
-\const(n\rho^{-1/4}(\rho^{1/4}\ell)^3)^{-2} (|d_1|+|d_2|).
\end{equation}
Here $d_1$ and $d_2$, depending on $\Psi$, are given as explained
in Lemma~\ref{local}. 
\end{lm}
\begin{proof}
As explained above we choose $M$ to be of order
$n\rho^{-1/4}(\rho^{1/4}\ell)^3$.
We then choose $\widetilde{\Psi}$ as explained above.
Then (\ref{eq:n+localization}) holds.
We also know that the possible $\hn_+$ values of $\widetilde{\Psi}$ range
in an interval of length $M$. We do not know however, where this
interval is located. The assumption (\ref{eq:Psiassumption}) will
allow us to say more about the location of the interval.

In fact, it follows from (\ref{eq:Psiassumption}),
(\ref{eq:n+localization}) that 
$\left(\widetilde{\Psi},H^n_{\ell,r,R}\widetilde{\Psi}\right)\leq 0$.
It is then a consequence of Lemma~\ref{lm:neutralcondensation} that 
$\left(\widetilde{\Psi},\hn_+\Psi\right)\leq 
\const n\rho^{-1/4}(\rho^{1/4}\ell)^3$. This of course establishes
that the allowed $\hn_+$ values are less than 
$\const' n\rho^{-1/4}(\rho^{1/4}\ell)^3$ for some constant $\const'>0$.
\end{proof}
Our final task in this section is to bound $d_1$ and $d_2$.  
We have that $d_1=(\Psi,H^n_{\ell,r,R}(1)\psi)$, 
where $H^n_{\ell,r,R}(1)$ is the part of the Hamiltonian 
$H^n_{\ell,r,R}$ containing all the terms with the coefficents
$\hw_{pq,\mu\nu}$ for which precisely one or three indices are 0. 
These are the terms bounded in Lemmas~\ref{lm:p000} and
\ref{lm:pqm0}. These lemmas are stated as one-sided bounds. 
It is clear from the proof that they could have been stated as 
two sided bounds. Alternatively we may observe that
$H^n_{\ell,r,R}(1)$ is unitarily equivalent to $-H^n_{\ell,r,R}(1)$. 
This follows by applying the unitary transform which maps 
all operators $a^*_p$ and $\an_p$ with $p\ne0$ to $-a^*_p$ and $-\an_p$.
{F}rom Lemmas~\ref{lm:p000} and \ref{lm:pqm0} we therefore immediately
get the following bound on $d_1$.
\begin{lm}[Control of $d_1$]\label{lm:d1}\hfill\\
With the notation above we have for all $\varepsilon>0$
\begin{eqnarray*}
  |d_1|&\leq& \varepsilon^{-1}{8\pi}\ell^{-3}R^2
  \left(\Psi,\hn_0\hn_+\Psi\right)+\varepsilon\left(\Psi,\left(\hn_+r^{-1}
      +\hw_{00,00}(\hn_0+1-\rho\ell^3)^2\right)\Psi\right)
  \\ &&{}
  +\varepsilon\left(\Psi,\sum_{p,m,p',m'\ne0}\hw_{mp',pm'}a^*_ma^*_{p'}\an_{m'}\an_{p}
    \Psi\right).
\end{eqnarray*}
\end{lm}
Likewise, we  have that $d_2=(\Psi,H^n_{\ell,r,R}(2)\psi)$, 
where $H^n_{\ell,r,R}(2)$ is the part of the Hamiltonian 
$H^n_{\ell,r,R}$ containing all the terms with precisely two $\an_0$
or two $a^*_0$. 
i.e., these are the terms in the Foldy Hamiltonian, which do not
commute with $\hn_+$. 
\begin{lm}[Control of $d_2$]\label{lm:d2}\hfill\\ 
There exists a constant $\const>0$ such that if
$(\rho^{1/4}\ell) t^3>\const$  and 
$(\rho^{1/4}\ell)\rho^{-1/12} $, $t$, and
$\omega(t)(\rho^{1/4}\ell)^{-1}$ are less than $\const^{-1}$ and
$\Psi$ is a wave function with 
$(\Psi,H^n_{\ell}\Psi)\leq 0$ then with the notation above we have
$$
|d_2|\leq \const \rho^{5/4}\ell^3(\rho^{1/4}\ell)
+4\pi\ell^{-3}R^2\left(\Psi,\hn_+\hn_0\Psi\right).
$$
\end{lm}
\begin{proof} If we replace all the operators  $a^*_p$ and $\an_p$ with
$p\ne0$ in the Foldy Hamiltonian by $-ia^*_p$ and $i\an_p$ we get a
unitarily equivalent operator. This operator however differs from the
Hamiltonian $H_{\rm Foldy}$ only by a change of sign on the part that
we denoted $H^n_{\ell,r,R}(2)$. Since both operators satisfy the bound 
in Corollary~\ref{cl:lbfh} we conclude that 
\begin{eqnarray*}
  |d_2|&\leq&
  \left(\Psi,\left[\mfr{1}{2}\gamma^{-1}\sum_p|p|^2a^*_p\an_p
      +\mfr{1}{2}\sum_{pq\ne0}\hw_{pq,00}\left(
        a^*_pa^*_0\an_0 \an_q+a^*_0a^*_p\an_q
        \an_0\right)\right]\Psi\right)
  \\ &&{}
  +\const n^{5/4}\ell^{-3/4}.
\end{eqnarray*}
Note that both sums above define positive operators. 
This is trivial for the first sum. For the second it follows from
(\ref{eq:p00p'}) in Lemma~\ref{lm:hw} since $a^*_0\an_0$ commutes
with all $a^*_p$ and $\an_p$ with $p\ne0$. The lemma now follows
from  (\ref{eq:p00p'}) and from Lemma~\ref{lm:neutralcondensation}.
\end{proof}

\section{Proof of Foldy's law}\label{sec:foldyslaw}

We first prove Foldy's law in a small cube. 
Let $\Psi$ be a normalized $n$-particle wave function. 
We shall prove that with an appropriate choice of $\ell$
\begin{equation}\label{eq:foldysmall}
\left(\Psi,H^n_\ell\Psi\right)\geq\left(\mfr{4\pi}{3}\right)^{1/3}A \rho\ell^3
\left(\rho^{1/4}+o\left(\rho^{1/4}\right)\right)
\end{equation}
where $A$ is given in (\ref{int}). Note that $A<0$. 
It then follows from Lemma~\ref{lm:reducing} that
$$
 E_0\geq 
 (1+L/\ell)^3\gamma \left(\mfr{4\pi}{3}\right)^{1/3}A \rho\ell^3
\left(\rho^{1/4}+o\left(\rho^{1/4}\right)\right)
  -\const (L/\ell)^2\rho^2 \ell^5
  -\frac{\omega(t)N}{2\ell}.
$$
Thus, since $N=\rho L^3$ we have 
$$
\lim_{L\to\infty}\frac{E_0}{N}\geq \gamma 
\left(\mfr{4\pi}{3}\right)^{1/3}A 
\left(\rho^{1/4}+o\left(\rho^{1/4}\right)\right)
  -\const\rho^{1/4} \omega(t)\left(\rho^{1/4}\ell\right)^{-1}.
$$
Foldy's law (\ref{gse}) follows since we shall choose (see below)
$t$ and $\ell$ in such a way 
that as $\rho\to \infty$ we have $t\to0$ and hence $\gamma\to1$ and 
$\omega(t)(\rho^{1/4}\ell)^{-1}\to0$ (see condition (\ref{cond1})
below).

It remains to prove (\ref{eq:foldysmall}).  
First we fix the long and short distance potential cutoffs
\begin{equation}\label{eq:rRchoice}
 R=\omega(t)^{-1}\ell,\quad\hbox{and}\quad 
r=\rho^{-3/8}(\rho^{1/4}\ell)^{-1/2}.
\end{equation}
We may of course assume that $\left(\Psi,H^n_\ell\Psi\right)\leq0$.
Thus $n$ satisfies the bound in Lemma~\ref{lm:neutralcondensation}.
We proceed in two steps. In Lemma~\ref{lm:case1} Foldy's law in the
small boxes  is proved
under the restrictive assumption given in (\ref{eq:case1}) below.
Finally, in Theorem~\ref{thm:foldyslaw} Foldy's law in the small boxes
is proved by considering the alternative case that (\ref{eq:case1})
fails.  Let us note that, logically speaking, this could have been
done in the reverse order. I.e., we could, instead, have begun with
the case that (\ref{eq:case1}) fails.
At the end of the section we combine Theorem~\ref{thm:foldyslaw}
with Lemma~\ref{lm:reducing} to show that Foldy's law in the
small box implies Foldy's law Theorem~\ref{thm:foldy}.

At the end of this section we show how to choose
$\ell$ and $t$ so that Theorem~\ref{thm:foldyslaw} implies 
(\ref{eq:foldysmall}) and hence Theorem~\ref{thm:foldy}, 
as explained above.

\begin{lm}[Foldy's law for $H^n_\ell$: restricted version]\label{lm:case1}
Let $R$ and $r$ be given by (\ref{eq:rRchoice}).
There exists a constant $\const>0$ such that if
$(\rho^{1/4}\ell) t^3>C$  and 
$(\rho^{1/4}\ell)\rho^{-1/12} $, $t$, and
$\omega(t)(\rho^{1/4}\ell)^{-1}$ are less than $\const^{-1}$
then, whenever
\begin{eqnarray}
  \lefteqn{n\ell^{-3}R^2\left(\Psi,\hn_+\Psi\right)}&&\label{eq:case1}\\
  &\leq& \const^{-1}\left(\Psi,\left(\hw_{00,00}(\hn_0-\rho\ell^3)^2+
      \sum_{p,m,p',m'\ne0}\hw_{mp',pm'}a^*_ma^*_{p'}\an_{m'}\an_{p}
    \right)\Psi\right),\nonumber
\end{eqnarray}
we have that
\begin{eqnarray*}
  \left(\Psi,H^n_\ell\Psi\right)
  \geq -In^{5/4}\ell^{-3/4}
  &-\const \rho^{5/4}\ell^3& \Bigl(\omega(t)(\rho^{1/4}\ell)^{-1}+
  \omega(t)^{-2}\rho^{-1/8}(\rho^{1/4}\ell)^{13/2}\\
  &&{}+
  +\rho^{-1/8}(\rho^{1/4}\ell)^{7/2}\Bigr),
\end{eqnarray*}
with $I$ as in Lemma \ref{lm:lbqh}. 
\end{lm}
\begin{proof}We assume $\left(\Psi,H^n_\ell\Psi\right)\leq0$.
We proceed as in the beginning of Sect.~\ref{sec:nn+}, but we now
use (\ref{eq:lbfh1}) of Corollary~\ref{cl:lbfh} instead of
(\ref{eq:lbfh2}). We then get
\begin{eqnarray*}
  H^n_\ell&\geq& 
  -In^{5/4}\ell^{-3/4}
  -4\pi n^{5/4}\ell^{-3/4}(n\ell)^{-1/4}
  -\mfr{1}{2}nR^{-1}-\const n\rho r^2 \\&&
  +\mfr{1}{2}\hw_{00,00}\left[\left(\hn_0-\rho\ell^3\right)^2
    -\hn_0\right]
  \\&&
  -4\pi[\rho-n\ell^{-3}]_+\hn_+R^2
  -4\pi\hn_+^2\ell^{-3}R^2
  \\&&
  -\varepsilon^{-1}{8\pi}\ell^{-3}R^2 \hn_0\hn_+
  -\varepsilon\hw_{00,00}
  (\hn_0+1-\rho\ell^3)^2
  \\&& 
  -\varepsilon\hn_+r^{-1} 
  +(\mfr{1}{2}-\varepsilon)\sum_{p,m,p',m'\ne0}
  \hw_{mp',pm'}a^*_ma^*_{p'}\an_{m'}\an_{p}.
\end{eqnarray*}
If we now use the assumption (\ref{eq:case1}) and the facts that
$\hn_+\leq n$, $\hn_0\leq n$, and $\hw_{00,00}\leq4\pi R^2\ell^{-3}$
we see with appropriate choices of $\varepsilon$ and $\const$ that 
\begin{eqnarray*}
  H^n_\ell&\geq& 
  -In^{5/4}\ell^{-3/4}
  -4\pi n^{5/4}\ell^{-3/4}(n\ell)^{-1/4}
  -\mfr{1}{2}nR^{-1}-\const n\rho r^2 -\const R^2\ell^{-3}(n+1) \\&&
  -\const R^2\ell^{-3}|n-\rho\ell^3|(\hn_++1)
  -\const\hn_+r^{-1}.
\end{eqnarray*}
If we finally insert the choices of $R$ and $r$ and use
Lemma~\ref{lm:neutralcondensation} we arrive at the bound in the
lemma.
\end{proof}

\begin{thm}[Foldy's law for $H^n_\ell$]\label{thm:foldyslaw}\hfill\\
There exists a $\const>0$ such that if
$(\rho^{1/4}\ell) t^3>\const$  and 
$(\rho^{1/4}\ell)\rho^{-1/12} $, $t$, and
$\omega(t)(\rho^{1/4}\ell)^{-1}$ are less than $\const^{-1}$
then for any normalized $n$-particle wave function $\Psi$ we have 
\begin{eqnarray}
  \left(\Psi,H^n_\ell\Psi\right)
  \geq -In^{5/4}\ell^{-3/4}
  &-\const \rho^{5/4}\ell^3& \Bigl(\omega(t)(\rho^{1/4}\ell)^{-1}
  +\omega(t)^{-1}\rho^{-1/16}(\rho^{1/4}\ell)^{29/4}\nonumber\\
  &&{}
  +\rho^{-1/8}(\rho^{1/4}\ell)^{7/2}\Bigr)\label{eq:finalestimate},
\end{eqnarray}
where $I$ is defined in Lemma~\ref{lm:lbqh} 
with $r$ and $R$ as in (\ref{eq:rRchoice}).
\end{thm}
\begin{proof}
According to Lemma~\ref{lm:case1} we may assume that
\begin{eqnarray}
  \lefteqn{n\ell^{-3}R^2\left(\Psi,\hn_+\Psi\right)}&&\label{eq:case1op}\\
  &\geq& \const^{-1}\left(\Psi,\left(\hw_{00,00}(\hn_0-\rho\ell^3)^2+
      \sum_{p,m,p',m'\ne0}\hw_{mp',pm'}a^*_ma^*_{p'}\an_{m'}\an_{p}
    \right)\Psi\right),\nonumber
\end{eqnarray}
where $\const$ is at least as big as the constant in Lemma~\ref{lm:case1}. 
We still assume that
$\left(\Psi,H^n_\ell\Psi\right)\leq0$. 

We begin by bounding $d_1$ and $d_2$ using 
Lemmas~\ref{lm:d1} and \ref{lm:d2}. We have from 
Lemmas~\ref{lm:neutralcondensation} and \ref{lm:d2} that 
\begin{eqnarray*}
|d_2|&\leq& \const \rho^{5/4}\ell^3(\rho^{1/4}\ell)
+\const \ell^{-1}\omega(t)^{-2} n^2
\rho^{-1/4}(\rho^{1/4}\ell)^3\\
&\leq&
\const[n\rho^{-1/4}(\rho^{1/4}\ell)^3]^2\rho^{5/4}\ell^3
\left((\rho^{1/4}\ell)^{-11}+\omega(t)^{-2}(\rho^{1/4}\ell)^{-7}\right)
\\&\leq&
\const[n\rho^{-1/4}(\rho^{1/4}\ell)^3]^2\rho^{5/4}\ell^3
\omega(t)^{-2}(\rho^{1/4}\ell)^{-7}.
\end{eqnarray*}
In order to bound $d_1$ we shall use (\ref{eq:case1op}). 
Together with Lemma~\ref{lm:d1} this gives (choosing $\varepsilon=1/2$ 
say)
\begin{eqnarray*}
  |d_1|&\leq& \const\ell^{-3}R^2n
  \left(\Psi,\hn_+\Psi\right)+\mfr{1}{2}\left(\Psi,\left(\hn_+r^{-1}
      +\hw_{00,00}(n-\rho\ell^3+1)\right)\Psi\right).
\end{eqnarray*}
Inserting the choices for $r$ and $R$ and using Lemma~\ref{lm:neutralcondensation}
gives 
$$
|d_1|\leq \const[n\rho^{-1/4}(\rho^{1/4}\ell)^3]^2\rho^{5/4}\ell^3
\left(\omega(t)^{-2}(\rho^{1/4}\ell)^{-7}+\rho^{-1/8}(\rho^{1/4}\ell)^{-17/2}\right)
$$
where we have also used that we may assume that
$\rho^{-1/8}(\rho^{1/4}\ell)^{-9/2}$ is small.
The assumption (\ref{eq:Psiassumption}) now reads 
$$
\left(\Psi,H^n_{\ell,r,R}\Psi\right)\leq
-\const \rho^{5/4}\ell^3
\left(\omega(t)^{-2}(\rho^{1/4}\ell)^{-7}
  +\rho^{-1/8}(\rho^{1/4}\ell)^{-17/2}\right).
$$
If this is not satisfied we see immediately 
that the bound (\ref{eq:finalestimate}) holds.

Thus from Lemma~\ref{lm:n+localization} it follows that we can find 
a normalized $n$-particle wavefunction $\widetilde{\Psi}$ 
with 
\begin{equation}\label{eq:n+^2expectation}
\left(\widetilde{\Psi},\hn_+\widetilde{\Psi}\right)
\leq \const n\rho^{-1/4}(\rho^{1/4}\ell)^3
\quad\hbox{and}\quad\left(\widetilde{\Psi},\hn_+^2\widetilde{\Psi}\right)
\leq \const n^2\rho^{-1/2}(\rho^{1/4}\ell)^6
\end{equation}
such that   
$$
\left(\Psi,H^n_{\ell,r,R}\Psi\right)\geq
\left(\widetilde{\Psi},H^n_{\ell,r,R}\widetilde{\Psi}\right)
-\const \rho^{5/4}\ell^3
\left(\omega(t)^{-2}(\rho^{1/4}\ell)^{-7}
  +\rho^{-1/8}(\rho^{1/4}\ell)^{-17/2}\right).
$$

In order to analyze
$\left(\widetilde{\Psi},H^n_{\ell,r,R}\widetilde{\Psi}\right)$
we proceed as in the beginning of Sect.~\ref{sec:nn+}.
This time we use Lemmas 
\ref{lm:cutoffs},\ref{lm:0000},\ref{lm:p0q0},\ref{lm:p000}, and
\ref{lm:pqm0}
together with Lemma~\ref{lm:lbqh} instead of Corollary~\ref{cl:lbfh}.
We obtain
\begin{eqnarray*}
  H^n_{\ell,r,R}&\geq& \mfr{1}{2}\hw_{00,00}\left[\left(n-\rho\ell^3\right)^2
    +\left(\hn_+\right)^2-2 \left(n-\rho\ell^3\right)\hn_+
    -\hn_0\right]
  \\&&
  -4\pi[\rho-n\ell^{-3}]_+\hn_+R^2-4\pi\hn_+^2\ell^{-3}R^2
  -\varepsilon\hn_+r^{-1} 
  -\varepsilon^{-1}{8\pi} \ell^{-3}R^2\hn_0\hn_+ 
  \\&&-\varepsilon\hw_{00,00}(\hn_+-1)^2+(\mfr{1}{2}-\varepsilon)\sum_{p,m,p',m'\ne0}
  \hw_{mp',pm'}a^*_ma^*_{p'}\an_{m'}\an_{p}
  \\&&
  -\mfr{1}{2}\left(n-\rho\ell^3\right)^2\hw_{00,00}
  -4\pi n^{5/4}\ell^{-3/4}(n\ell)^{-1/4}-In^{5/4}\ell^{-3/4}.
\end{eqnarray*}
This time we shall however not choose $\varepsilon$ small, but rather big.
Note that since $w_{r,R}(x,y)\leq r^{-1}$ we have 
$
\sum\limits_{p,m,p',m'\ne0}\hw_{mp',pm'}a^*_ma^*_{p'}\an_{m'}\an_{p}\leq
r^{-1}\hn_+(\hn_+-1)
$, 
which follows immediately from
\begin{eqnarray*}
  \lefteqn{\sum_{p,m,p',m'\ne0}
    \hw_{mp',pm'}a^*_ma^*_{p'}\an_{m'}\an_{p}}&&\\
  &=&\iint w_{r,R}(x,y)
  \left(\sum_{p,m\ne0}u_m(x)u_{p}(y)\an_m\an_p\right)^*
  \sum_{p,m\ne0}u_m(x)u_{p}(y)\an_m\an_p\,dx\,dy.
\end{eqnarray*}
We therefore have
\begin{eqnarray*}
  H^n_{\ell,r,R}&\geq& -In^{5/4}\ell^{-3/4}-4\pi
  n^{5/4}\ell^{-3/4}(n\ell)^{-1/4}
  -\const R^2\ell^{-3}\hn_0
  \\&&
  -\const\ell^{-3}R^2|\rho\ell^3-n|\hn_+-4\pi\hn_+^2\ell^{-3}R^2
  -\varepsilon\hn_+r^{-1} 
  -\varepsilon^{-1}{8\pi} \ell^{-3}R^2\hn_0\hn_+ 
  \\&&-\varepsilon\const R^2\ell^{-3}\hn_+^2-\varepsilon\hn_+^2r^{-1}.
\end{eqnarray*}
If we now insert the choices of $r$ and $R$,
take the expectation in the state given by $\widetilde\Psi$, and
use (\ref{eq:n+^2expectation}) and  the bound on $n$ 
from Lemma~\ref{lm:neutralcondensation}
we arrive at
\begin{eqnarray*}
  \left(\widetilde{\Psi},H^n_{\ell,r,R}\widetilde{\Psi}\right)&\geq& 
  -In^{5/4}\ell^{-3/4}-\const \rho^{5/4}\ell^3
  \Bigl[(\rho^{1/4}\ell)^{-1} +\omega(t)^{-2}(\rho^{1/4}\ell)^{-1}
  \\&&+\omega(t)^{-2}\rho^{-1/8}(\rho^{1/4}\ell)^{11/2}
  +\omega(t)^{-2}\rho^{-1/4}(\rho^{1/4}\ell)^{8}
  +\varepsilon \rho^{-1/8}(\rho^{1/4}\ell)^{7/2}
  \\&&
  +\varepsilon^{-1}\omega(t)^{-2}(\rho^{1/4}\ell)^{5}
  +\varepsilon \omega(t)^{-2}\rho^{-1/4}(\rho^{1/4}\ell)^{8}
  +\varepsilon \rho^{-1/8}(\rho^{1/4}\ell)^{19/2} \Bigr].
\end{eqnarray*}
If we now choose
$\varepsilon=\omega(t)^{-1}\rho^{1/16}(\rho^{1/4}\ell)^{-9/4}$
we arrive at (\ref{eq:finalestimate}).
\end{proof}

\begin{proof}[Completion of the proof of Foldy's law,
  Theorem~\ref{thm:foldy}]
We have accumulated various errors and we want to show that 
they can all be made small. There are basically two parameters that
can be adjusted, $\ell$ and $t$. Instead of $\ell$ it is convenient to
use
$
   X=\rho^{1/4}\ell
$. We shall choose $X$ as a function of $\rho$ such that 
$X\to\infty$ as $\rho\to\infty$. 
{F}rom Lemma~\ref{lm:sbn} we know that for some fixed $C>0$
$C^{-1}\rho\ell^3\leq n\leq C\rho\ell^3$.
Hence according to 
(\ref{eq:paramlimits}) with $r$ and $R$ given in 
(\ref{eq:rRchoice})
we have that $I\to -\left(\mfr{4\pi}{3}\right)^{1/3}A$
as $\rho\to\infty$ if 
\begin{eqnarray}
 \omega(t)^{-1}X&\to&\infty\label{cond1}\\
 \rho^{1/4}X&\to&\infty\label{cond2}\\
 t^3X&\to&\infty\label{cond3}\\
 t&\to&0\label{cond4}.
\end{eqnarray}
The hypotheses of
Theorem~\ref{thm:foldyslaw} are valid if 
(\ref{cond1}),(\ref{cond3}),
(\ref{cond4}), and
\begin{eqnarray}
  \rho^{-1/12}X&\to&0\label{cond6}
\end{eqnarray}
hold.
{F}rom Lemma~\ref{lm:neutralcondensation}, for which the hypotheses are now
automatically satisfied, we have that
$
 n=\rho\ell^3(1+O(\rho^{-1/8}X^{3/2})
$
and from (\ref{cond6}) we see that $n$ is $\rho\ell^3$ to leading
order. 

With these conditions we find that the first term 
on the right side of (\ref{eq:finalestimate}) is, in the limit
$\rho\to\infty$, exactly Foldy's law. 
The conditions that the other terms in (\ref{eq:finalestimate})
are of lower order are 
\begin{eqnarray}
  (X/\omega(t))^{4/25}\rho^{-1/100}X&\to&0\label{cond7}\\
  \rho^{-1/28}X&\to&0\label{cond8}
\end{eqnarray}
together with (\ref{cond1}).

It remains to show that  we can satisfy the conditions
(\ref{cond1}--\ref{cond8}). 
Condition (\ref{cond2}) is trivially satisfied since both $\rho$ and
$X$ tend to infinity. 
Since $\omega(t)\sim t^{-4}$ for small $t$ we see that 
(\ref{cond3}) is implied by (\ref{cond1}).
Condition (\ref{cond6}) is implied by (\ref{cond8}), which is 
in turn implied by (\ref{cond1}) and (\ref{cond7}). 
The remaining two conditions (\ref{cond1}) and (\ref{cond7}) are
easily satisfied by an approriate choice of $X$ and $t$ as functions
for $\rho$ with $X\to\infty$ and $t\to0$ as $\rho\to\infty$.   
In fact, we simply need $\rho^{1/116}t^{-16/29}\gg X\gg t^{-4}$.

The bound (\ref{eq:foldysmall}) has now been established. Hence
Foldy's law Theorem~\ref{thm:foldy} follows as discussed in the 
beginning of the section. 
\end{proof}
\appendix

\section{Appendix: Localization of large matrices}\label{app:local}
\bigskip

The following theorem allows us to reduce a big Hermitean matrix,
${\cal A}$, to a smaller principal submatrix without changing the lowest
eigenvalue very much.   ( The $k^{\rm th}$ supra- (resp. infra-)  diagonal
of a matrix ${\cal A}$ is the submatrix consisting of all elements $a_{i,
i+k}$ (resp. $a_{i+k , i}$). )

\begin{thm}[Localization of large matrices]\label{local} 
  Suppose that ${\cal A}$ is an $N\times N$ Hermitean matrix and let
  ${\cal A}^k$, with $k=0,1,...,N-1$, denote the matrix consisting of
  the $k^{\rm th}$ supra- and infra-diagonal of ${\cal A}$.  Let $\psi
  \in {\bf C}^N$ be a normalized vector and set $d_k = (\psi , {\cal
    A}^k \psi) $ and $\lambda = (\psi , {\cal A} \psi) =
  \sum_{k=0}^{N-1} d_k$.  \ ($\psi$ need not be an eigenvector of
  ${\cal A}$.) \ 
  
  Choose some positive integer $M \leq N$.  Then, with $M$ fixed,
  there is some $n \in [0, N-M]$ and some normalized vector $ \phi \in
  {\bf C}^N$ with the property that $\phi_j =0$ unless $n+1 \leq j
  \leq n+M$ \ (i.e., $\phi $ has length $M$) and such that
  \begin{equation}\label{localerror}
    (\phi , {\cal A} \phi) \leq \lambda + \frac{C}{ M^2}
    \sum_{k=1}^{M-1} k^2 |d_k|
    +C\sum_{k=M}^{N-1} |d_k|\ ,
  \end{equation}
  where $C>0 $ is a  universal constant. (Note that the first sum starts
  with $k=1$.)
\end{thm}

\begin{proof} It is convenient to  extend the matrix ${\cal A}_{i,j}$ to all
  $-\infty < i,j < +\infty$ by defining ${{\cal A}}_{i,j} =0$
  unless $1\leq i,j \leq N$. Similarly, we extend the vector
  $\psi$ and we define the numbers $d_k$ and the matrix ${\cal A}^k$
  to be zero when $k\not\in [0,N-1]$. 
  We shall give the construction for $M$
  odd, the $M$ even case being similar. 

  For $s \in {\bf Z}$ set $f(s ) = A_M [M+1-2|s |]$  if $2|s |
  < M $ and $f(s ) =0$ otherwise. Thus, $f(s ) \not= 0$ for precisely $M$
  values of $s $. Also, $f(s )=f(-s )$. $A_M$ is chosen so that
  $\sum_s  f(s )^2 =1$.

  For each $m \in {\bf Z}$ define the vector 
  $\phi^{(m)}$ by  $\phi^{(m)}_j = f(j-m){\psi}_j$.
  We then define $K^{(m)} =(\phi^{(m)}, {{\cal A}} \phi^{(m)}) -(\lambda
  +\sigma) (\phi^{(m)}, \phi^{(m)})$. (The number $\sigma$ will be chosen
  later.) 
  After this, we define $K=\sum_m K^{(m)} $. Using the fact that
  $\sum_s   f(s )^2 =1$, we have that
  \begin{eqnarray*}
    \sum_m
    (\phi^{(m)},{{\cal A}}\phi^{(m)})&=&
    \sum_m
    \sum_{k=0}(\phi^{(m)},{{\cal A}}^k\phi^{(m)})
    =\sum_s \sum_{k}f(s)f(k+s)(\psi,{{\cal A}}^k\psi)\\
    &=&\sum_s \sum_{k=0}f(s)f(k+s){d}_k
  \end{eqnarray*}
  and
  \begin{equation}
    \lambda = \lambda \sum_m
    (\phi^{(m)},\phi^{(m)})
    =\sum_s\sum_{k=0}f(s)^2(\psi,{{\cal A}}^
    k\psi)
    =\sum_s\sum_{k} f(s)^2{d}_k\  
  \end{equation}

  Hence
  \begin{equation}
    K=\sum_m K^{(m)} = -\sigma- \sum_{k=1}^{N-1} d_k \gamma_k 
  \end{equation}
  with
  \begin{equation}\label{gamma}
    \gamma_k = \frac{1}{2} \sum_s  \left[ f(s ) - f(s +k) \right]^2 \ .
  \end{equation}

  Let us choose $\sigma = -\sum_{k=1}^{N-1} d_k \gamma_k$. Then,
  $\sum_m K^{(m)} =0$. Recalling that not all of the $\phi^{(m)}$ equal
  zero, we conclude that there is at least one value of $m$ such that 
  (i) $\phi^{(m)}\not= 0$ and (ii) 
  $(\phi^{(m)}, {{\cal A}} \phi^{(m)}) \leq (\lambda
  +\sigma) (\phi^{(m)}, \phi^{(m)})$.  

  This concludes the proof of (\ref{localerror}) except for showing
  that 
  $  \gamma_k \leq C \frac{k^2}{k^2+M^2}$ for all $M$ and $k$. 
  This is evident from the easily computable large $M$ asymptotics
  in (\ref{gamma}).

\end{proof}

\section{Appendix: A double commutator bound}\label{app:double}

\begin{lm}\label{lm:double} Let $-\Delta_N$ be the Neumann Laplacian 
  of some bounded open set ${\cal O}$. 
  Given $\theta\in C^\infty(\overline{\cal O})$ 
  with $\supp |\nabla\theta|\subset {\cal O}$ satisfying
  $\|\partial_i\theta\|\leq \const t^{-1}$, 
  $\|\partial_i\partial_j\theta\|\leq \const t^{-2}$, 
  $\|\partial_i\partial_j\partial_k\theta\|\leq \const t^{-3}$, 
  for some $0<t$ and all $i,j,k=1,2,3$. 
  Then for all $s>0$ we have the operator inequality
  \begin{equation}\label{eq:doublecom2}
    \left[\left[
        \frac{\left(-\Delta_N\right)^2}{-\Delta_N+s^{-2}},
        \theta\right]
      ,\theta\right]\geq-\const t^{-2}\frac{-\Delta_N}{-\Delta_N+s^{-2}}
    -\const s^2t^{-4}. 
  \end{equation}
  We also have the norm bound
  \begin{equation}\label{eq:doublecom1}
    \left\|\left[\left[
          \frac{-\Delta_N}{-\Delta_N+s^{-2}},
          \theta\right]
        ,\theta\right]\right\|\leq\const (s^2t^{-2}+s^4t^{-4}). 
  \end{equation}
\end{lm}
\begin{proof}
  We calculate the commutator 
  \begin{eqnarray*}
    \left[\frac{\left(-\Delta_N\right)^2}{-\Delta_N+s^{-2}},
      \theta\right]&=&
    s^{-2}\frac{1}{-\Delta_N+s^{-2}}
    \left[-\Delta_N,\theta\right]
    \frac{1}{-\Delta_N+s^{-2}}\left(-\Delta_N\right)
    \\&&{}
    +\frac{-\Delta_N}{-\Delta_N+s^{-2}}\left[-\Delta_N,\theta\right].
  \end{eqnarray*}
  Likewise we calculate the double commutator
  \begin{eqnarray}
    \lefteqn{
      \left[\left[\frac{\left(-\Delta_N\right)^2}{-\Delta_N+s^{-2}},
          \theta\right],\theta\right]=
      -\frac{-\Delta_N}{-\Delta_N+s^{-2}}
      \left[\left[-\Delta_N,\theta\right]\theta\right]
      \frac{-\Delta_N}{-\Delta_N+s^{-2}}}\nonumber
    &&\\{}
    &&+\left[\left[-\Delta_N,\theta\right]\theta\right]
    \frac{-\Delta_N}{-\Delta_N+s^{-2}}
    +\frac{-\Delta_N}{-\Delta_N+s^{-2}}
    \left[\left[-\Delta_N,\theta\right]\theta\right]\nonumber\\
    &&{}-2s^{-4}\frac{1}{-\Delta_N+s^{-2}}
    \left[-\Delta_N,\theta\right]
    \frac{1}{-\Delta_N+s^{-2}}
    \left[\theta,-\Delta_N\right]
    \frac{1}{-\Delta_N+s^{-2}}.\label{eq:doublecom}
  \end{eqnarray}
  Note that $\left[\left[-\Delta_N,\theta\right]\theta\right]
  =-2\left(\nabla\theta\right)^2$ and thus the first term above is
  positive. 

  We claim that 
  \begin{equation}\label{eq:comsquare}
    \left[-\Delta_N,\theta\right]\left[\theta,-\Delta_N\right]
    \leq -\const t^{-2}\Delta_N+\const t^{-4}. 
  \end{equation}
  To see this we simply calculate
  $$
  \left[-\Delta_N,\theta\right]\left[\theta,-\Delta_N\right]
  =-\sum_{i,j}^3\left( 
    4\partial_i (\partial_i\theta)(\partial_j\theta)\partial_j
    +(\partial_i^2\theta)(\partial_j^2\theta)
    +2(\partial_i\theta)(\partial_i\partial_j^2\theta)\right)
  $$
  The last two terms are bounded by $\const  t^{-4}$. 
  For the first term we have by the Cauchy-Schwarz inequality
  for operators, 
  $BA^*+AB^*\leq\varepsilon^{-1}AA^*+\varepsilon BB^*$, for all 
  $\varepsilon>0$,  that
  $$
  -\sum_{i,j}^3\partial_i (\partial_i\theta)(\partial_j\theta)\partial_j=
  \sum_{i,j}^3\left(\partial_i (\partial_i\theta)\right)
  \left(\partial_j(\partial_j\theta)\right)^*
  \leq -3\sum_{i}^3\partial_i (\partial_i\theta)
  (\partial_i\theta)\partial_i
  $$
  and this is bounded above by $-3t^{-2}\Delta_N$ and we get 
  (\ref{eq:comsquare}). 
  Inserting (\ref{eq:comsquare}) into (\ref{eq:doublecom}), 
  recalling that the first term is positive, we obtain
  \begin{eqnarray*}
    \left[\left[\frac{\left(-\Delta_N\right)^2}{-\Delta_N+s^{-2}},
        \theta\right],\theta\right]
    &\geq& -2(\nabla\theta)^2\frac{-\Delta_N}{-\Delta_N+s^{-2}}
    -2\frac{-\Delta_N}{-\Delta_N+s^{-2}}(\nabla\theta)^2\\&&
    -\const t^{-2}\frac{-\Delta_N}{-\Delta_N+s^{-2}}
    -\const s^2 t^{-4}.
  \end{eqnarray*}
  Again using the Cauchy-Schwarz inequality,
  we have
  \begin{eqnarray*}
    \lefteqn{2(\nabla\theta)^2\frac{-\Delta_N}{-\Delta_N+s^{-2}}+
      2\frac{-\Delta_N}{-\Delta_N+s^{-2}}(\nabla\theta)^2}&&\\
    &\leq& 
    2t^{-2}\left(\frac{-\Delta_N}{-\Delta_N+s^{-2}}\right)^{1/2}
    \left(\nabla\theta\right)^4
    \left(\frac{-\Delta_N}{-\Delta_N+s^{-2}}\right)^{1/2}
    +2t^{-2}\left(\frac{-\Delta_N}{-\Delta_N+s^{-2}}\right)\\
    &\leq&\const t^{-2}\frac{-\Delta_N}{-\Delta_N+s^{-2}},
  \end{eqnarray*}
  and (\ref{eq:doublecom2}) follows. 

  The bound (\ref{eq:doublecom1}) is proved in the same
  way. Indeed,
  \begin{eqnarray*} 
    \lefteqn{\left[\left[\frac{-\Delta_N}{-\Delta_N+s^{-2}},
        \theta\right],\theta\right]=
    -s^{-2}\frac{1}{-\Delta_N+s^{-2}}[[-\Delta_N,\theta],\theta]
    \frac{1}{-\Delta_N+s^{-2}}}\hspace{2cm}&&\\
    &&{}+2s^{-2}\frac{1}{-\Delta_N+s^{-2}}
    \left[-\Delta_N,\theta\right]
    \frac{1}{-\Delta_N+s^{-2}}
    \left[\theta-\Delta_N\right]
    \frac{1}{-\Delta_N+s^{-2}},
  \end{eqnarray*}
and (\ref{eq:doublecom1}) follows from 
$\left[\left[-\Delta_N,\theta\right]\theta\right]
=-2\left(\nabla\theta\right)^2$ and (\ref{eq:comsquare}). 
\end{proof}

\end{document}